\documentclass[]{aastex631}
\submitjournal{MAPS}

\shorttitle{Winchcombe Fireball}
\shortauthors{McMullan et al.}
\graphicspath{{./}{figs/}}
\usepackage{todonotes}
\usepackage{soul}
\usepackage{verbatim}
\begin{document}

\title{The Winchcombe Fireball---that Lucky Survivor}

%% LaTeX will automatically break titles if they run longer than
%% one line. However, you may use \\ to force a line break if
%% you desire. In v6.31 you can include a footnote in the title.

\newcommand{\contrib}[1]{\textcolor{teal}{#1}}

\newcommand{\numb}[1]{\textcolor{orange}{#1}}
\newcommand{\source}[1]{\textsuperscript{\textcolor{blue}{[citation needed]}}}
\newcommand{\missnumber}{\numb{[NUMBER]}}
\newcommand{\checknumber}{\numb{[check number]}}

\newcommand{\densunitSI}{kg\,m$^{-3}$}
\newcommand{\kms}{km\,s$^{-1}$}
\newcommand{\ms}{m\,s$^{-1}$}
\newcommand{\kmssqu}{km\,s$^{-2}$}
\newcommand{\ablaunit}{kg\,MJ$^{-1}$}

\newcommand{\fireballTimeZero}{2021-02-28T21:54:17}
\newcommand{\codename}{DN210228\_02}
\newcommand{\meteorite}{\textit{Winchcombe}}

\newcommand{\DFN}{Desert Fireball Network}
\newcommand{\GFO}{Global Fireball Observatory}

\newcommand{\curtin}{School of Earth and Planetary Sciences, Curtin University, Perth WA 6845, Australia}
\newcommand{\uwo}{Department of Physics and Astronomy, University of Western Ontario, London, Ontario, N6A 3K7, Canada}
\newcommand{\imperial}{Impact and Astromaterials Research Centre, Dept. Earth Science and Eng., Imperial College London, SW7 2BP, UK}
\newcommand{\glasgow}{School of Geographical and Earth Sciences, University of Glasgow, Glasgow, UK}
\newcommand{\surec}{SUREC, University of Glasgow, Glasgow, UK}
\newcommand{\nhm}{Planetary Materials Group, Department of Earth Sciences, Natural History Museum, Cromwell Road, London SW7 5BD, UK}
\newcommand{\imcce}{IMCCE, Observatoire de Paris, PSL Research University, CNRS, Sorbonne Universités, UPMC Univ. Paris 06, Univ. Lille, 77 Av. Denfert-Rochereau, 75014, Paris, France}
\newcommand{\ou}{School of Physical Sciences, The Open University, Walton Hall, Milton Keynes, MK7 6AA, UK}
\newcommand{\oxford}{Department of Materials, University of Oxford, Oxford, UK}
\newcommand{\cambridge}{University of Cambridge, Cavendish Laboratory, Cambridge, UK}
\newcommand{\manchester}{Department for Earth and Environmental Sciences, The University of Manchester, Manchester, UK}

\newcommand{\ukfall}{UK Fireball Alliance (UKFAll)}
\newcommand{\ukfn}{UK Fireball Network (UKFN)}
\newcommand{\gfo}{Global Fireball Observatory (GFO)}
\newcommand{\gmn}{Global Meteor Network (GMN)}
\newcommand{\scamp}{System for the Capture of Asteroid and Meteorite Paths (SCAMP)}

\correspondingauthor{Sarah McMullan}
\email{s.mcmullan16@imperial.ac.uk }

\author[0000-0002-7194-6317]{Sarah McMullan}
\affiliation{\imperial}
\affiliation{\ukfn}
\affiliation{\ukfall}
\affiliation{\gfo}

\author[0000-0003-4166-8704]{Denis Vida}
\affiliation{\uwo}
\affiliation{\gmn}
\affiliation{\gfo}

\author[0000-0001-9226-1870]{Hadrien A. R. Devillepoix}
\affiliation{\curtin}
\affiliation{\gfo}

\author{Jim Rowe}
\affiliation{\ukfall}
\affiliation{\scamp}

% UKFALL
\author[0000-0002-7150-4092]{Luke Daly}
\affiliation{\glasgow}
\affiliation{\oxford}
\affiliation{\ukfn}
\affiliation{\ukfall}
\affiliation{\gfo}

\author[0000-0001-6113-5417]{Ashley J. King}
\affiliation{\nhm}
\affiliation{\ukfall}

% GFO
\author[0000-0003-2193-0867]{Martin Cup\'ak}
\affiliation{\curtin}
\affiliation{\gfo}

\author[0000-0002-5864-105X]{Robert M. Howie}
\affiliation{\curtin}
\affiliation{\gfo}

\author[0000-0003-2702-673X]{Eleanor K. Sansom}
\affiliation{\curtin}
\affiliation{\gfo}

\author[0000-0003-4766-2098]{Patrick Shober}
\affiliation{\imcce}
\affiliation{\gfo}

\author[0000-0002-8240-4150]{Martin C. Towner}
\affiliation{\curtin}
\affiliation{\gfo}

\author[0000-0002-8914-3264]{Seamus Anderson}
\affiliation{\curtin}
\affiliation{\gfo}

% UWO
\author{Luke McFadden}
\affiliation{\uwo}

% SCAMP
\newcommand{\cardiff}{National Museum Cardiff, Cardiff, C10 3NP, UK}
\author{Jana Hor\'ak}
\affiliation{\cardiff}
\affiliation{\scamp}
\affiliation{\ukfall}

\author[0000-0001-7137-6628]{Andrew R. D. Smedley}
\affiliation{\manchester}
\affiliation{\scamp}
\affiliation{\ukfall}

\author[0000-0003-4992-8750]{Katherine H. Joy}
\affiliation{\manchester}
\affiliation{\scamp}
\affiliation{\ukfall}

\author{Alan Shuttleworth}
\affiliation{\scamp}
\affiliation{\fripon}

% FRIPON
\newcommand{\fripon}{Fireball Recovery and InterPlanetary Recovery (FRIPON); France}
\author[0000-0002-5859-6586]{Francois Colas}
\affiliation{\imcce}
\affiliation{\fripon}

\author{Brigitte Zanda}
\affiliation{\fripon}

% UKFN
\author[0000-0002-2591-7902]{Áine C. O'Brien}
\affiliation{\glasgow}
\affiliation{\surec}
\affiliation{\ukfn}
\affiliation{\ukfall}

\author{Ian McMullan}
\affiliation{\ukfn}

\author{Clive Shaw}
\affiliation{\ukfn}
\affiliation{\cambridge}

\author{Adam Suttle}
\affiliation{\ukfn}

\author[0000-0001-7165-2215]{Martin D. Suttle}
\affiliation{\ukfn}
\affiliation{\ou}

\author{John S. Young}
\affiliation{\ukfn}
\affiliation{\cambridge}

% UKMON
\newcommand{\ukmon}{UK Meteor Observation Network (UKMON); UK}
\author{Peter Campbell-Burns}
\affiliation{\ukmon}
\affiliation{\ukfall}

\author{Richard Kacerek}
\affiliation{\ukmon}
\affiliation{\ukfall}

\author{Richard Bassom}
\affiliation{\ukmon}
\affiliation{\gmn}

\author{Steve Bosley}
\affiliation{\ukmon}

\author{Richard Fleet}
\affiliation{\ukmon}

\author{Dave Jones}
\affiliation{\ukmon}

\author[0000-0002-5769-4280]{Mark McIntyre}
\affiliation{\ukmon}
\affiliation{\gmn}

% NEMETODE
\newcommand{\nemetode}{NEMETODE Network; UK}

\author{Nick James}
\affiliation{\nemetode}

\author{Derek Robson}
\affiliation{\ukmon}
\affiliation{\nemetode}

% GMN
\author{Paul Dickinson}
\affiliation{\gmn}

\author[0000-0002-4681-7898]{Philip A. Bland}
\affiliation{\curtin}
\affiliation{\gfo}

\author[0000-0000-0000-0000]{Gareth S. Collins}
\affiliation{\imperial}
\affiliation{\ukfn}
\affiliation{\gfo}

% Sarah McMullan, Denis Vida, Hadrien A. R. Devillepoix, Jim Rowe, Luke Daly, Ashley J. King, Martin Cup\'ak, Robert M. Howie, Eleanor K. Sansom, Patrick Shober, Martin C. Towner, Seamus Anderson, Luke McFadden, Jana Hor\'ak, Andrew R. D. Smedley, Katherine H. Joy, Alan Shuttleworth, Francois Colas, Brigitte Zanda, Áine C. O'Brien, Ian McMullan, Clive Shaw, Adam Suttle, Martin D. Suttle, John S. Young, Peter Campbell-Burns, Richard Kacerek, Richard Bassom, Steve Bosley, Richard Fleet, Dave Jones, Mark McIntyre, Nick James, Derek Robson, Paul Dickinson, Philip A. Bland, Gareth S. Collins

%% Note that the \and command from previous versions of AASTeX is now
%% depreciated in this version as it is no longer necessary. AASTeX 
%% automatically takes care of all commas and "and"s between authors names.

%% AASTeX 6.31 has the new \collaboration and \nocollaboration commands to
%% provide the collaboration status of a group of authors. These commands 
%% can be used either before or after the list of corresponding authors. The
%% argument for \collaboration is the collaboration identifier. Authors are
%% encouraged to surround collaboration identifiers with ()s. The 
%% \nocollaboration command takes no argument and exists to indicate that
%% the nearby authors are not part of surrounding collaborations.

%% Mark off the abstract in the ``abstract'' environment. 
\begin{abstract}
%% IMPORTANT NOTE - Denis: I had to drop some details, as we are limited to only 200 words. I only tried to hit the most important takeaways
%Understanding how meteoroids transit the upper atmosphere in fireball events is critical to evaluating the likelihood a meteorite survived to the ground, then accurately calculating its fall position and orbit. 
On February 28, 2021, a fireball dropped $\sim0.6$~kg of recovered CM2 carbonaceous chondrite meteorites in South-West England near the town of Winchcombe.
We reconstruct the fireball's atmospheric trajectory, light curve, fragmentation behaviour, and pre-atmospheric orbit from optical records contributed by five networks.
The progenitor meteoroid was three orders of magnitude less massive ($\sim13$~kg) than any previously observed carbonaceous fall. The Winchcombe meteorite survived entry because it was exposed to a very low peak atmospheric dynamic pressure ($\sim0.6$~MPa) due to a fortuitous combination of entry parameters, notably low velocity (13.9~\kms{}). A near-catastrophic fragmentation at $\sim0.07$~MPa points to the body's fragility. 
Low entry speeds which cause low peak dynamic pressures are likely necessary conditions for a small carbonaceous meteoroid to survive atmospheric entry, strongly constraining the radiant direction to the general antapex direction. 
Orbital integrations show that the meteoroid was injected into the near-Earth region $\sim0.08$~Myr ago and it never had a perihelion distance smaller than $\sim0.7$~AU, while other CM2 meteorites with known orbits approached the Sun closer ($\sim0.5$~AU) and were heated to at least 100~K higher temperatures.
%In addition, a sudden flight vector change was observed after a fragmentation event towards the end of the bright flight, with the physical cause unexplained.
%The collaborative work from different optical observation networks, aided by a standardised astrometric data format, was instrumental to quickly calculate the meteorite fall area for Winchcombe, and will serve as a template for future meteorite fall events.
\end{abstract}

\section{Introduction}

On the 28th February 2021, at 21:54:16 UTC, a bright fireball lasting 8 seconds was observed above southern Wales, ending around Gloucester, UK \citep{king_science_winchcombe}. It was witnessed by over 1000 people and captured by many doorbell and dashboard cameras. It was also captured by 16 dedicated meteor/fireball cameras of the UK Fireball Alliance (UKFAll), making it a meteorite fall with one of the highest number of instrumental records to date. 

The UKFAll consortium was established in 2018 as a collaboration between the five meteor camera networks in the UK, with an aim to streamline data sharing and meteorite recovery efforts \citep{2020EPSC...14..705D}. The precursory work that UKFAll had done prior to this event enabled the team to share data, establish an initial strewn field, and handle press inquiries, all within 12 hours of the fall. This streamlined process enabled the recovery of a portion of the 339~g main mass the morning following the fall. The meteorite was discovered as a rubble pile on a driveway in the town of Winchcombe, about 60~km south of Birmingham \citep{king_science_winchcombe}. The rest of the main mass was collected from this same site the next day, with another 283~g of fragments recovered from the surrounding area over the next week \citep{metbull_110}. The meteorite was identified as a CM2 carbonaceous chondrite \citep{krot2014classification, suttle2021aqueous}; a rare type as only $\sim4$\% of meteorite falls globally are carbonaceous chondrites \citep{scott2014chondrites}.

Instrumentally observing a meteorite fall enables the computation of its pre-atmospheric orbit \citep{ceplecha1961multiple, 2020P&SS..19105036D}. This can link the meteorite sample to a particular source region in the solar system \citep{granvik2018identification}. Pairing an orbit with meteorite laboratory analyses enlightens our understanding of the composition of that particular region. Due to the presence of aqueous alteration \citep{bischoff1998aqueous}, carbonaceous chondrites are known to have formed close to the snow line in the outer solar system \citep{krot2015sources}. The exact mechanism of delivery of carbonaceous material into the asteroid belt is still a matter of discussion, but a migration event of the giant planets appears to be a necessary condition \citep{meech2020origins, vida2022direct}.

Carbonaceous chondrites contain abundant water and organic matter \citep{trigo2019accretion, matlovivc2022hydrogen}. They are proposed as a significant source of Earth's water and organic material so may be key to understanding the origins of life on Earth \citep{pizzarello2006nature, marty2012origins, alexander2012provenances}. Prior to the Winchcombe meteorite, only four carbonaceous chondrites have been linked to their pre-atmospheric orbit---Tagish Lake \citep{2000Sci...290..320B}, Maribo \citep{haack2012maribo}, Sutter's Mill \citep{2012Sci...338.1583J}, and Flensburg \citep{2021M&PS...56..425B}. However, none of them were well observed instrumentally from multiple stations and for most, trajectories were reconstructed from either casual images and videos or non-optical recordings.

Each of these events had a limited number of high-precision observations, especially from nearby dedicated cameras. Additionally, three out of four were daylight fireballs which makes calibration of video records more difficult, requiring the use of proxy objects as calibration points instead of stars \citep{borovivcka2014analysis}.

Winchcombe is the first carbonaceous chondrite fall which was recorded by multiple dedicated meteor/fireball cameras within 150~km of the fireball, and thus, observed with unprecedented detail. It was an evening event, providing both an opportunity for capturing high-precision optical recordings, as well as being widely witnessed---over 1000 eyewitness reports---resulting in much public interest.

In this paper, we perform a complete analysis of the fireball from the available optical records. In Section \ref{sec:data_methods} we describe the observations and camera networks that observed the fireball. In Section \ref{sec:traj} we discuss the trajectory and fragmentation modelling, continuing with the orbital analysis in Section \ref{sec:orbit}. We compare the strewn field calculation with the locations of the meteorites found in Section \ref{sec:darkflight}. Finally, in Section \ref{sec:discussion} we put Winchcombe in context with previous orbital carbonaceous chondrites and discuss the relevance of this unique fall.

\section{Data \& Methods} \label{sec:data_methods}

\begin{table}[!h]
    \caption{List of coordinates of UKFAll cameras that observed the Winchcombe fireball. The coordinates have been truncated to three decimal places due to privacy reasons, but the full trajectory solutions were produced using at least 5 decimal places ($\sim1$~m accuracy). Astrometry/Photometry indicates whether the data from the camera was used for astrometric picks (A) or for photometry (P). The altitude is given in the mean sea level convention, not WGS84. F is the plate scale, and A err is the astrometric fit error.
}
    \centering
    \begin{tabular}{ccccccccc}
    \hline 
    Location & Network & Latitude (\degr) & Longitude (\degr) & Alt. (m) & Instrument & Used for & F (arcmin/px) & A err (arcmin) \\
    \hline 
Cardiff      & SCAMP 	& 51.486 & -3.178 &  33 & all-sky video 	& A + P & 10.1		& 2.92		\\
Cambridge    & UKFN 	& 52.165 &  0.039 &   8 & all-sky photo	& -		& -		& -		\\
Chard        & UKMON 	& 50.878 & -2.950 & 100 & video			& A 	& 5.8		& 1.25		\\
Chelmsford   & NEMETODE & 51.745 &  0.494 &  45 & video			& - 	& -		& -		\\
Clanfield    & UKMON 	& 50.939 & -1.020 & 158 & video			& - 	& -		& -		\\
Honiton      & SCAMP 	& 50.802 & -3.184 & 119 & all-sky video 	& A + P & 10.1		& 3.42		\\
Hullavington & GMN 		& 51.535 & -2.149 & 103 & video 			& A 	& 3.8		& 0.83		\\
Lincoln      & UKFN		& 53.222 & -0.464 &  16 & all-sky photo 	& A 	& 2.0		& 0.64		\\
Loughborough & NEMETODE & 52.751 & -1.213 &  73 & video			& - 	& -		& -		\\
Manchester   & SCAMP 	& 53.474 & -2.234 &  69 & all-sky video 	& P 	& -		& -		\\
%Nuneaton     & AllSky7 	& 52.526389 & -1.454722 &  80 & video 			& - 	& -		& -		\\
Ringwood     & GMN 		& 50.858 & -1.778 &  24 & video 			& A 	& 3.8		& 1.32		\\
Tackley      & GMN 		& 51.883 & -1.306 &  80 & video 			& - 	& -		& -		\\
Welwyn       & UKFN 	& 51.268 & -0.394 &  78 & all-sky photo 	& A 	& 2.0		& 1.04		\\
Wilcot       & UKMON 	& 51.352 & -1.802 & 133 & video 			& P 	& -		& -		\\
%Antwerp (BE) & AllSky7 	& 51.213359 &  4.454574 &  16 & video 			& -		& -		& -		\\
    \hline 
    \end{tabular}
    
    \label{tab:stations}
\end{table}

\begin{figure}
    \centering
    \includegraphics[width=0.8\textwidth]{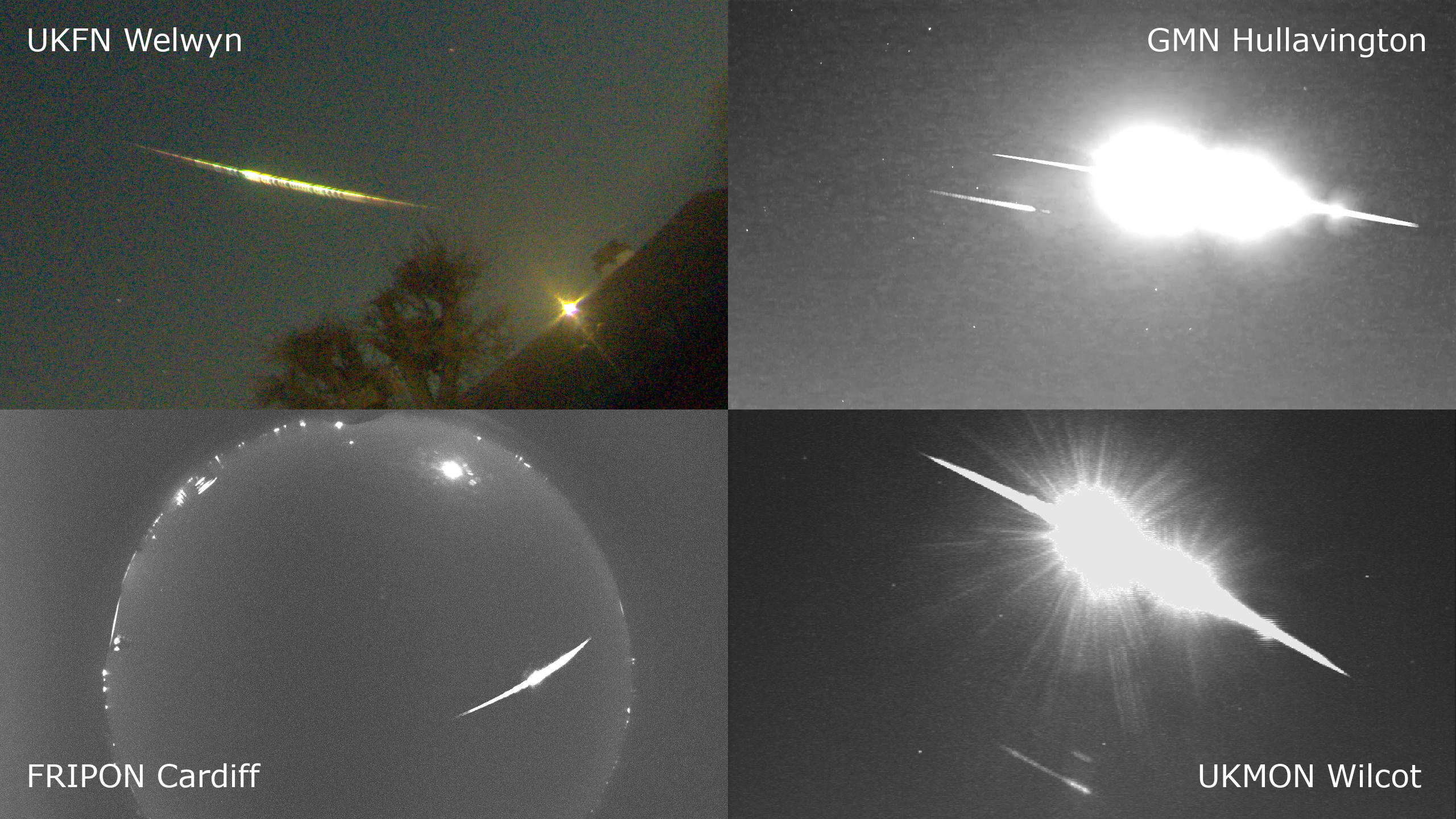}
    \caption{The Winchcombe fireball as seen from Welwyn (upper left), Hullavington (upper right), Cardiff (bottom left), and Wilcot (bottom right). The Hullavington and Wilcot images show a reflection parallel to the fireball. The fireball is moving from left to right in all images.}
    \label{fig:mosaic}
\end{figure}

The details of cameras that observed the fireball are given in Table \ref{tab:stations}. All networks which contributed optical observations are part of the \textit{UK Fireball Alliance}\footnote{\url{https://www.ukfall.org.uk/}} \citep{2020EPSC...14..705D}. Figure \ref{fig:mosaic} shows the image of the fireball from several cameras.

After the fall, all astrometry and photometry measurements produced by individual networks were mutually exchanged through the \textit{Global Fireball Exchange} format \citep{2020EPSC...14..856R}. Standard specifications are documented at \url{https://github.com/UKFAll/standard}, and the full final measurements are given in Supplementary Materials in this format.

Although each network employs its own reduction software suite for day-to-day operations, all astrometric and photometric data used for the analysis presented in this work have been re-measured following the methods of \citet{vida2021global}. In this section, we provide a brief description of each camera network.

\subsection{SCAMP/FRIPON}

SCAMP, the System for the Capture of Asteroid and Meteorite Paths, is the UK arm of the French Fireball Recovery and InterPlanetary Observation Network (FRIPON)\footnote{\url{https://www.fripon.org}} that extends over Europe \citep{2020A&A...644A..53C}. This network uses cameras with an all-sky lens to capture high-resolution video recordings (30 frames per second) of fireballs \citep{2015EPSC...10..800C}. There are currently 7~cameras in the UK network, with the aim to have 72~cameras in total to provide full coverage of the UK and Ireland.

\subsection{UKFN/GFO}

UKFN, the UK Fireball Network is the UK arm of the Global Fireball Observatory (GFO) collaboration\footnote{\url{https://gfo.rocks}}  \citep{2020P&SS..19105036D}.
This network uses all-sky cameras based on the DSLR system developed by the Desert Fireball Network (DFN) in Australia \citep{2017ExA....43..237H}.
These cameras capture a 27~s long-exposure photograph every 30~seconds. Absolute timing along the fireball track is encoded using a liquid crystal shutter and the de Bruijn method of \citet{2017M&PS...52.1669H}. The cameras require a spacing of $<$200~km for accurate fireball detection and observation \citep{devillepoix2019observation}; there are currently 6~cameras deployed in the UK, with plans to expand the network to a total of 11 cameras in the British Isles.

\subsection{GMN}

The Global Meteor Network (GMN)\footnote{\url{https://globalmeteornetwork.org}} operates over 700 video meteor stations in 38 countries \citep{vida2021global}. The stations use low-cost consumer-grade IMX291 and IMX307 CMOS sensors paired with wide-field lenses (most commonly $88^{\circ} \times 48^{\circ}$). All cameras are operated at 25 frames per second. 3.6~mm and 6~mm f/0.95 lenses are most commonly used, giving a similar field of view and sensitivity to a human observer (limiting magnitude $+6.0^{\mathrm{M}} \pm 0.5^{\mathrm{M}}$. The cameras are connected to Raspberry Pi single-board computers which run open-source software \citep{vida2016open, vida2018first}. Currently (circa. mid-2022), the GMN operates around 220 cameras in the UK.

\subsection{UKMON}

UKMON\footnote{\url{https://ukmeteornetwork.co.uk}}, the UK Meteor Network, is a group of amateur astronomers who use commercial CCTV video cameras for meteor monitoring \citep{2014JIMO...42..139C}. The network's main focus are fainter meteors and meteor showers, however the cameras also detect fireballs. The UKMON mainly uses GMN camera systems but also operates several older analog Watec cameras. UKMON currently has over 200 cameras in the UK, but at the time of the Winchcombe fall only around 30 cameras were installed.

\subsection{NEMETODE}

NEMETODE\footnote{\url{http://www.nemetode.org}}, the Network for Meteor Triangulation and Orbit Determination, is an amateur group with a network of analog and digital cameras to monitor the night sky for meteors and meteor showers, mostly based on UFOCapture software \citep{2013JIMO...41...84S}. Currently, NEMETODE has over 40 stations, generally with multiple cameras at each station, with significant coverage of much of Northern England and Ireland.

% \subsection{AllSky7}

% The European AllSky7 Network\footnote{\url{https://www.allsky7.net}} is a consortium of owners of AllSky7 camera systems. One system consists of seven video cameras positioned at different angles to achieve all-sky coverage. The systems are recording 24/7 and are able to detect daytime fireballs \citep{hankey2020all}. The cameras use 4~mm lenses which have the field of view and sensitivity similar to that of human eye (limiting magnitude of $\sim4^M$). The network currently has systems in 18 countries, with four stations in the UK.

\subsection{Other data}

Other observational data of the Winchcombe fall were collected along with the optical data described above. However, as they did not inform the astrometric and entry modelling of the Winchcombe fall we only briefly summarise them below:

\begin{itemize}
    \item A low-resolution visible spectrum was captured by two cameras (NEMETODE and UKMON).
    \item Some infrasound signals were detected by sensors from the Raspberry Shake \& Boom network \footnote{\url{https://raspberryshake.org}}, however no useful measurements could be made due to data timestamping issues.
    \item No seismic signals were detected by any seismographs within 200~km.
\end{itemize}

\section{Trajectory}\label{sec:traj}

In this section, we describe the details of the trajectory, discuss the data reduction procedure and calibration quality, and present the results of fireball ablation modelling.

\subsection{Astrometry and photometry}

All optical data sets were manually calibrated and reduced using the \texttt{SkyFit2} software \citep{vida2021global}\footnote{The code is available in the RMS repository: \url{https://github.com/CroatianMeteorNetwork/RMS}}. Both the all-sky and narrow-field data were calibrated using the radial distortion model with odd terms up to the seventh order ($r, r^3, r^5, r^7$), taking atmospheric refraction and lens anisotropy into account. All calibrations showed only random errors with no systematic trends. Table \ref{tab:stations} summarises the cameras used in the solution, together with the plate scales and the average astrometric fit errors, which were on the order of a few arc minutes for all systems. Of the total 16 cameras which observed the fireball, only seven were used in the final trajectory solution. Others were less optimal due to large distance, bad geometry, CCD blooming, or frame drops. We note that these data would also be useful if the seven picked stations did not offer the best view of the fireball.

The unsaturated light curve was exclusively measured on SCAMP video data for magnitudes fainter than $-8^{\mathrm{M}}$, at which point the cameras saturated (Cardiff, Honiton, and Manchester stations). All other cameras were either already saturated or observed the fireball through thin clouds which would degrade the quality of photometry measurements. Fortuitously, during the time of SCAMP camera saturation, the analogue CCD camera video from Wilcot showed an unsaturated lens reflection which was used to measure the brightest portion of the fireball. Independent absolute calibration of the reflection could not be done, the measurements yielding only an instrumental magnitude estimate. However, there were several common points with the unsaturated SCAMP portion of the light curve which were used to scale the instrumental magnitude of the reflection, allowing the full light curve to be reconstructed to a high degree of accuracy. Fig. \ref{fig:ablation_deceleration_model} shows the measured light curve.

\subsection{Atmospheric trajectory}

\begin{figure}
    \centering
    \includegraphics[width=0.8\textwidth]{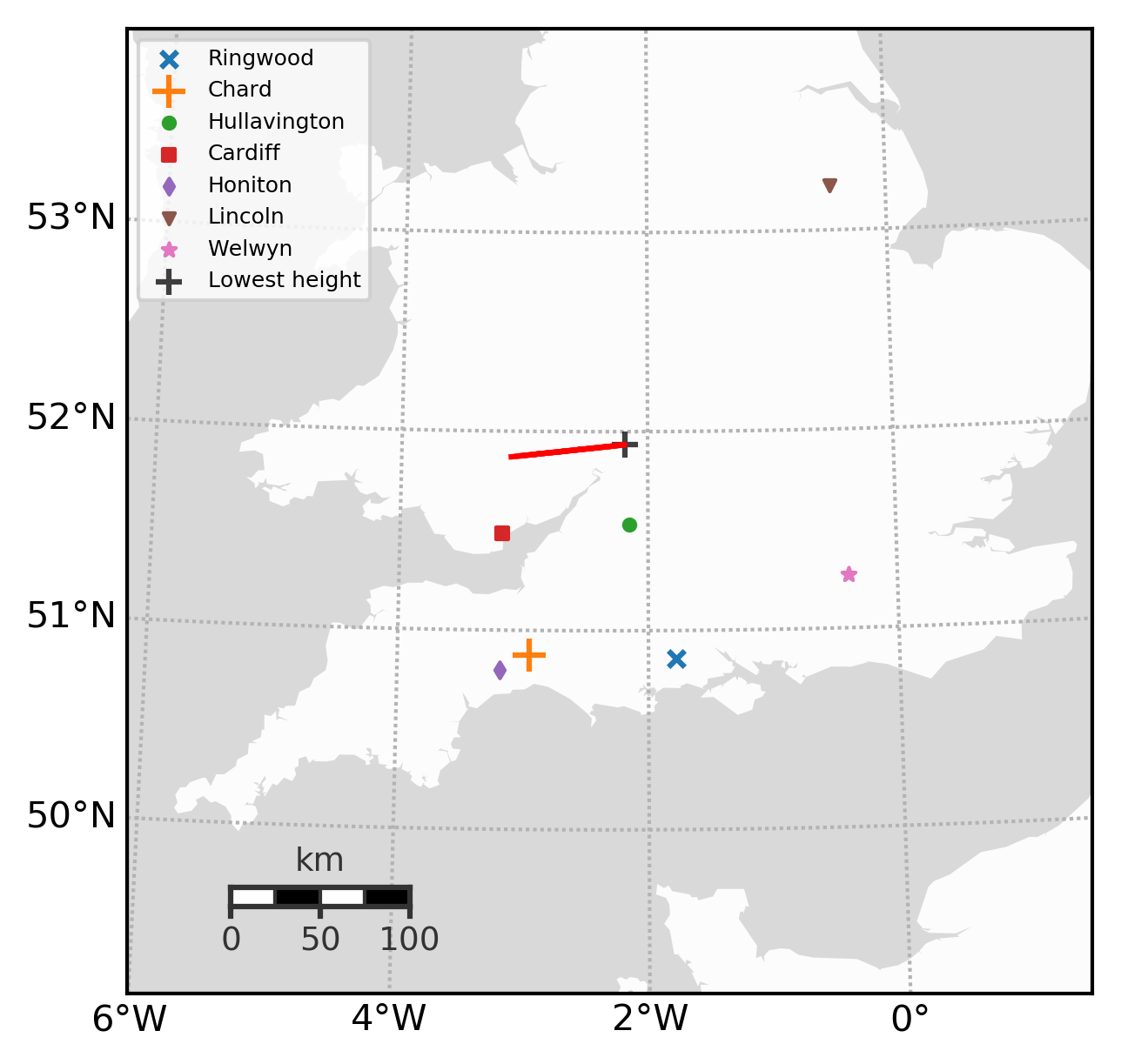}
    \caption{The Winchcombe fireball trajectory (red line) and locations of cameras used for astrometric measurements.}
    \label{fig:map}
\end{figure}

The nominal trajectory and orbital solution were calculated using the method of \citet{vida2020estimating}\footnote{The code is available in the WesternMeteorPyLib repository: \url{https://github.com/wmpg/WesternMeteorPyLib}}. The uncertainties were computed by adding Gaussian noise that is two times larger than the measured random fit errors, as per \cite{vida2020estimating}, and re-fitting the trajectory solution. The trajectory details are given in Table \ref{tab:traj_sum}. Figure~\ref{fig:map} shows the fireball trajectory in relation to the seven stations. All selected stations are within 200~km of the fireball, and the Hullavington station was only 50~km away, allowing it to capture the details of fragmentation and track the final fragments just before the dark flight began. Despite most stations being south of the fireball, the observation geometry was favourable --- the maximum convergence angle of 89$^{\circ}$ was between the Cardiff and Welwyn stations.

\begin{table*}
\centering
\caption{Fireball trajectory parameters with associated $1\sigma$ uncertainties. All values are given in the Earth-fixed system. The values of initial and final azimuth and altitude differ because the trajectory of the fireball is considered to be a straight line in the Earth-Centred Inertial (ECI) coordinate system, and are recomputed to ground-fixed values to be compatible with previous publications. The final azimuth and altitude do not take the deviating fragment into account.}
\label{tab:traj_sum}
\begin{tabular}{lrr}
& Beginning & End\\
\hline\hline 
Time              & 21:54:15.88         & 21:54:24.12 \\
Latitude (+N)     & 51.870970$^{\circ}$ & 51.940114$^{\circ}$ \\
                  & $\pm$15.3~m         & $\pm$16.2~m \\
Longitude (+E)    & -3.109378$^{\circ}$ & -2.096335$^{\circ}$ \\
                  & $\pm$8.8~m          & $\pm$5.6~m \\
Height (km)       & 90.599              & 27.554 \\
                  & $\pm$0.020          & $\pm$0.015 \\
Velocity (\kms{}) & 13.86               & $\sim 3$ \\
                  & $\pm 0.01$          & - \\
Azimuth           & 263.342$^{\circ}$   & 263.906$^{\circ}$ \\
                  & 0.046$^{\circ}$     & - \\
Altitude          & 41.919$^{\circ}$    & 41.530$^{\circ}$ \\
                  & 0.029$^{\circ}$     & - \\
\end{tabular}
\end{table*}

\begin{figure}
    \centering
    \includegraphics[width=0.6\textwidth]{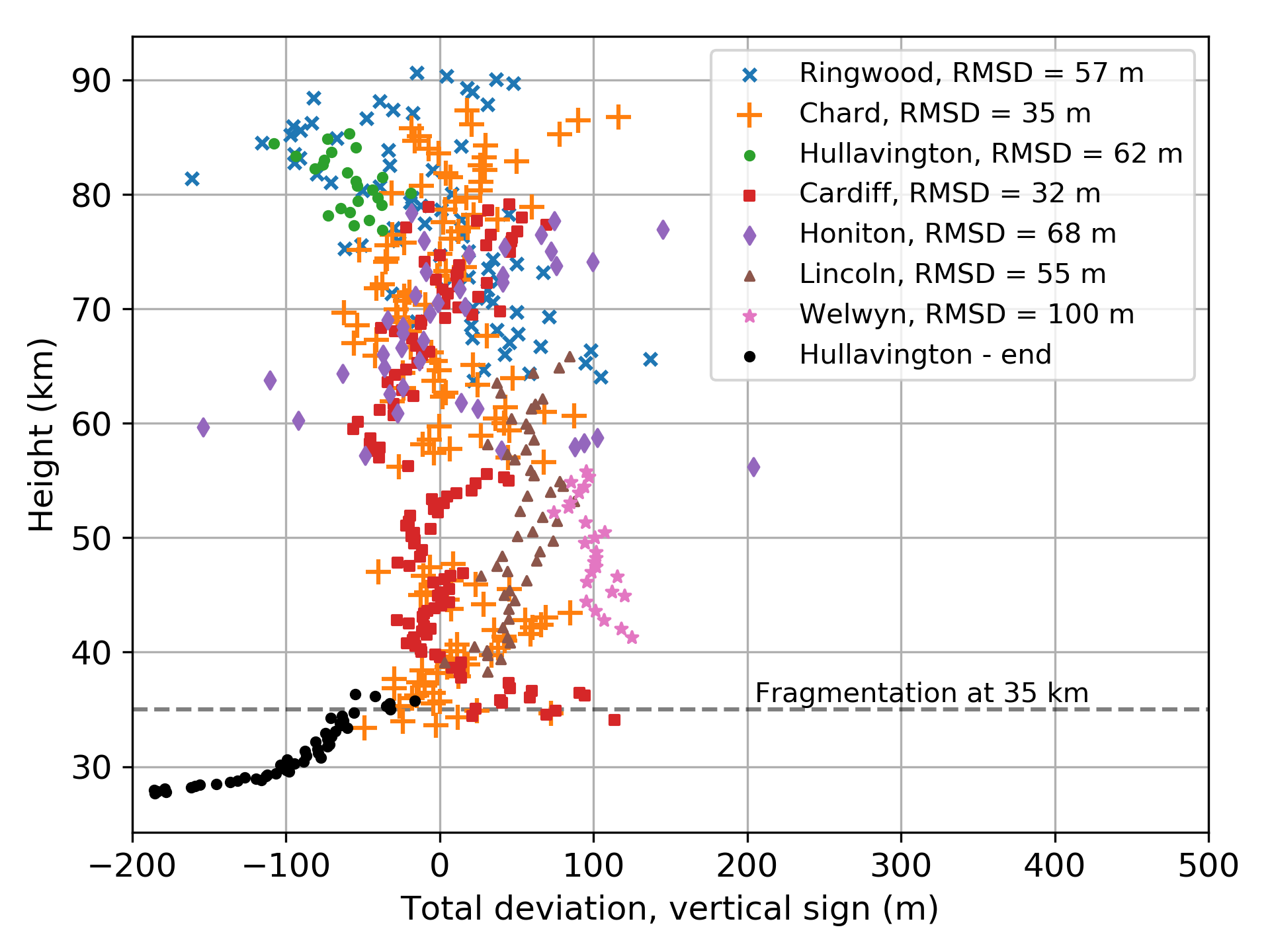}
    \caption{Total trajectory fit residuals versus height (the sign of the deviation is taken from the vertical direction). RMSD stands for root-mean-square deviation. The measurements from Hullavington were only done at the beginning and the end, as the fireball was too bright in the middle of the flight for accurate astrometric picks.}
    \label{fig:traj_residuals}
\end{figure}

Figure \ref{fig:traj_residuals} shows the trajectory fit residuals from a straight line. The trajectory fit residuals were all below 100~m across the observed span of almost 100~km. Following \citet{vida2020estimating}, the initial velocity was computed as the average velocity up to the time when deceleration became statistically significant. This is achieved by progressively including more points from the beginning of the trajectory until the end in a linear time vs. distance fit, and choosing the solution with the smallest standard deviation.
The fireball was first observed at the height of 90.6~km moving at a velocity of 13.86~\kms{}, and it was last observed at 27.6~km decelerating below the ablation limit at 3~\kms{}.

The Winchcombe meteoroid experienced several major fragmentation, dramatically increasing the observed fireball brightness and deceleration. The trajectory followed a straight line up until the final fragmentation at a height of 35~km. A sudden change in the direction of fragments was observed afterwards. The final portion of the trajectory which showed the deviation was only observed from the Hullavington station due to its closeness to the fireball and higher sensitivity. This final portion was either outside of the fields of view of other cameras, or they were not sensitive enough to observe it. The observed deviation was not due to calibration issues---the total observed deviation from a straight-line trajectory was 12 arc minutes, and the astrometry fit accuracy around the end of the fireball was 0.83 arc minutes. The apparent cross-track velocity of the fragment relative to a straight line was $\sim100$~\ms{}; however, this represents a lower limit as the orientation of the plane of fragmentation cannot be measured from a single-station observation. Only the part of the trajectory above 35~km was used for orbit estimation, to avoid any influence of the deviating fragment on the radiant.

\subsection{Final mass estimation}

\begin{figure}
    \centering
    \includegraphics[width=0.8\textwidth]{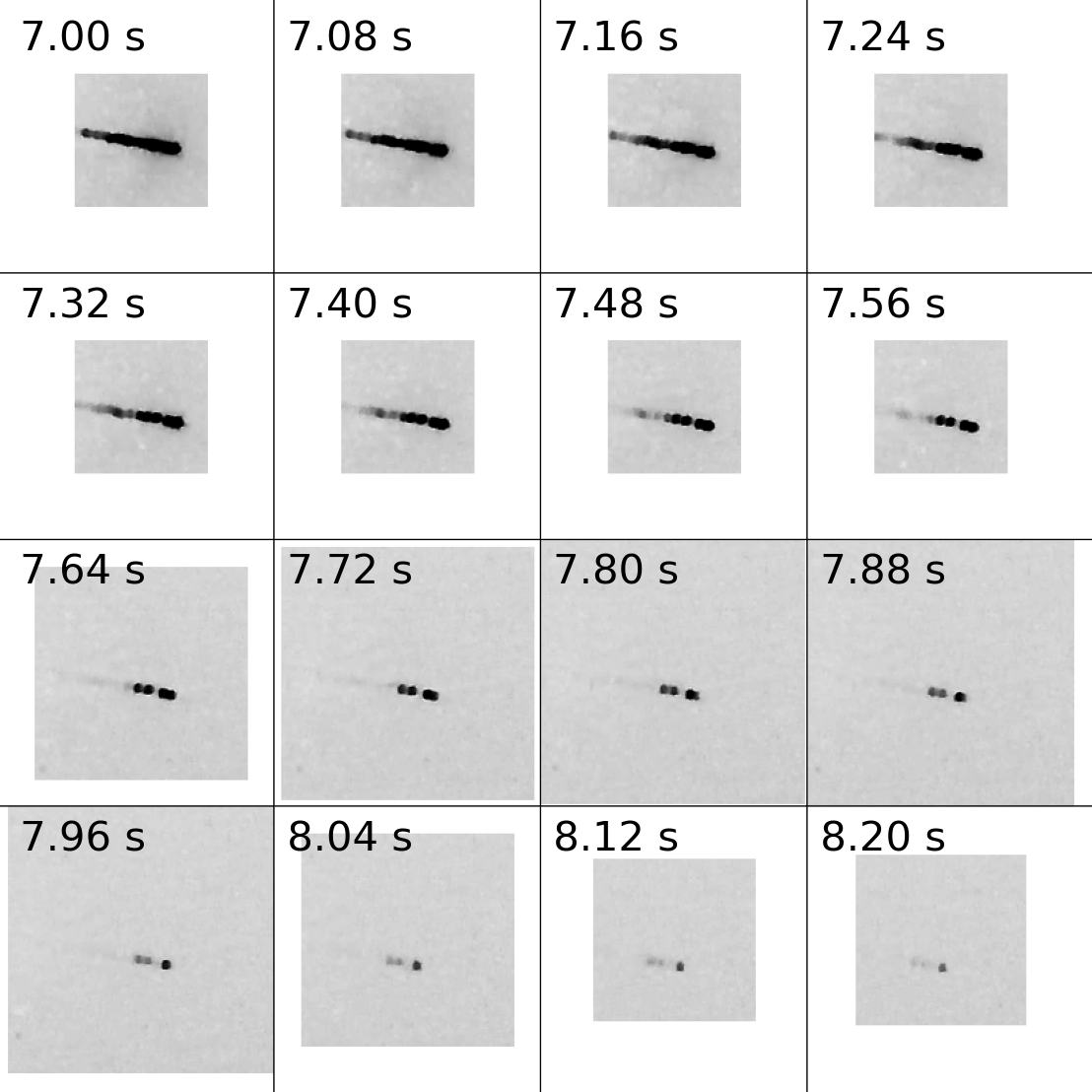}
    \caption{Colour-inverted mosaic of the last moments (every second frame) observed from Hullavington. Four discrete fragments can be seen at the end.}
    \label{fig:frags}
\end{figure}

\begin{figure}
    \centering
    \includegraphics[width=0.8\textwidth]{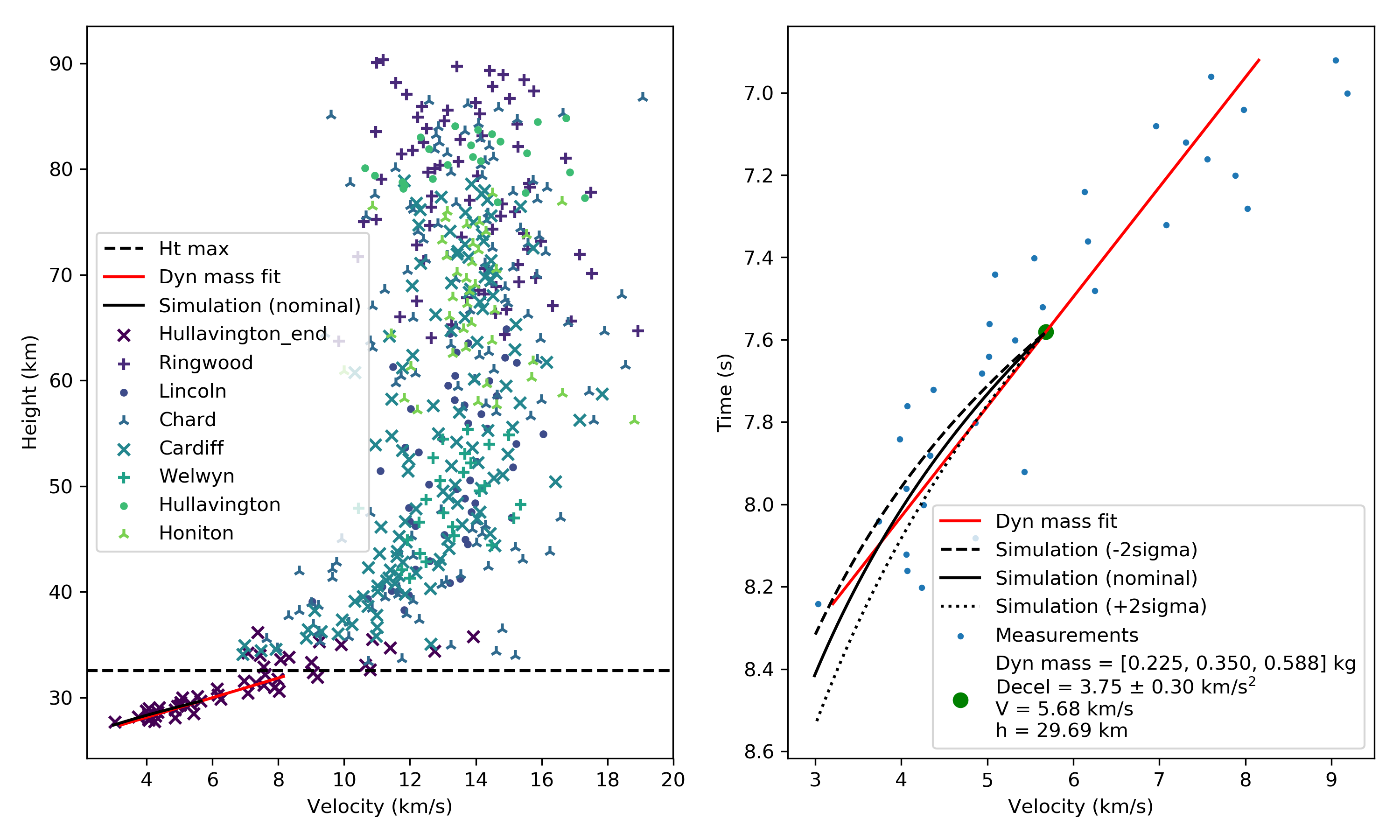}
    \caption{Left: Observed point-to-point velocities, and the dynamic mass fit (red line) done using final velocity measurements below the dashed line. Right: A section of velocity measurements at the end of the fireball (blue dots) used for the dynamic mass fit (red line). The mass was estimated in the middle (green dot). Black lines indicate the nominal simulation (solid) and the $2\sigma$ variations (dashed and dotted).}
    \label{fig:dyn_mass}
\end{figure}

This final fragmentation produced four discrete fragments (Figure \ref{fig:frags}) that could be individually tracked until they dimmed below magnitude $+2^{\mathrm{M}}$. These measurements allowed an accurate estimation of the dynamic mass of the largest fragment (Figure \ref{fig:dyn_mass}). Following a classical approach \citep{mccrosky1971lost}, a line was fit on time vs. velocity measurements near the fireball's end to obtain an estimate of the velocity and the deceleration. The dynamic mass $m_{dyn}$ is computed as:

\begin{equation}
    m_{dyn} = \frac{1}{\rho_m^2} \left ( \frac{\Gamma A \rho_a v^2}{\dot{v}} \right)^3 \,,
\end{equation}

\noindent where $\rho_m$ is the meteoroid bulk density, $\Gamma$ is the drag factor\footnote{$\Gamma$ is referred to as the drag factor in many meteoroid trajectory works, including \citep{Ceplecha2005}. The aerodynamic drag coefficient, $c_d$ = 2$\Gamma$ \citep{Bronshten1983, 2015aste.book..257B}.}, $A$ is the shape factor, and $\rho_a$ is the atmospheric mass density at the point where the velocity $v$ and deceleration $\dot{v}$ are measured.

In this method, the underlying assumption is that the mass loss is no longer the dominant driver of energy loss and that the fit can be fully described by the single-body drag equation. A bulk density of $\rho_m = $2100~\densunitSI{} was used, informed by the atmospheric ablation characteristics which indicate a carbonaceous body. This is consistent with the density of the recovered meteorites from micro-X-ray computed tomography ($\sim$2090~\densunitSI{} \citealt{king_science_winchcombe}). The product $\Gamma A$ can be treated as a free parameter, and has previously been found empirically to fall in the 0.5--1.0 range near the end of fireballs, representing spherical to cylindrical shapes \citep{borovicka2003moravka, borovivcka2015instrumentally, gritsevich2014first}.

% The product $\Gamma A$ of $\sim$0.5 is typical of more spherical shapes ($A=1.21$, $\Gamma\sim1$) while a product of 1.0 is closer to cylindrical shapes ($A=1.4$, $\Gamma\sim1.3$).

To constrain the mass by adjusting the $\Gamma A$ factor, and to compute the final location of the fragments prior to dark flight, we introduce a novel method. We perform a forward integration of single-body ablation equations \citep{ceplecha1998meteor} starting at the point where the dynamic mass is estimated. The equations are integrated until the velocity falls below 3~\kms{}, i.e. the ablation limit. An intrinsic ablation coefficient of $\sigma = 0.005$~kg~MJ$^{-1}$ is used \citep{ceplecha1998meteor}. $\Gamma A$ factor is adjusted until the simulated velocity matches the observed velocity at the end. The final geographical coordinates and height above ground are computed at the cessation of ablation and are used for dark flight modelling. The azimuth and elevation of the trajectory with respect to the ground at this endpoint are computed taking the drop due to gravity and the curvature of the Earth into account.

$\Gamma A = 0.55$ was found to best fit the observations. A range of masses between 0.225 and 0.588~kg was measured (95\% confidence interval) for the main fragment, with 0.350~kg being the nominal value. The dynamic mass was estimated at the height of 29.69~km and at the velocity of 5.68~\kms{}, at which a deceleration of $3.75\pm0.30$~\kmssqu{} was measured. The NRLMSISE-00 model \citep{picone2002nrlmsise} was used to obtain the atmospheric mass density. After integrating the ablation equations, only a further 5\% of mass was lost (the final nominal value for the main fragment was 0.330~kg). This nominal estimate for the lead fragment is consistent with the largest sample recovered which has a mass of 0.319~kg. The dynamic mass estimates for the smaller fragments are between 50 and 100~g, which is also consistent with the other recovered samples \citep{king_science_winchcombe}.

\subsection{Dynamics and Fragmentation Modelling} 

The dynamics, light curve, and fragmentation behaviour were modelled using the \cite{borovivcka2013kovsice} semi-empirical ablation model. The manual modelling procedure described in detail in \cite{borovivcka2020two} was followed. The luminous efficiency model used was from the same paper. Of the many semi-analytical approaches developed to model meteoroid trajectories and fragmentation processes in our atmosphere \cite[e.g.][]{johnston2018radiative, wheeler2017fragment}, this method gives us the highest fidelity and enables direct modelling of each observed feature and fragment individually.

In summary, the initial meteoroid is modelled as a classical single body that fragments at manually determined points. Physical parameters of the initial meteoroid and individual fragments are also manually estimated. All generated fragments are considered single bodies. The fragmentation points are informed by brightness increases in the light curve (i.e. flares) and increases in the observed deceleration. Most fragmentations were modelled as a release of an eroding fragment \citep[see ][ for more details]{borovivcka2015instrumentally} which is a body that rapidly erodes by the release of mm-sized grains. The grain masses are distributed according to a power law (differential mass index of $s = 2.0$ is assumed) within a given range of masses. The amount of erosion is regulated by the erosion coefficient $\eta$, which determines how much mass is eroded from the fragment per unit of kinetic energy loss.

\begin{figure}
    \centering
    \includegraphics[width=1.\textwidth]{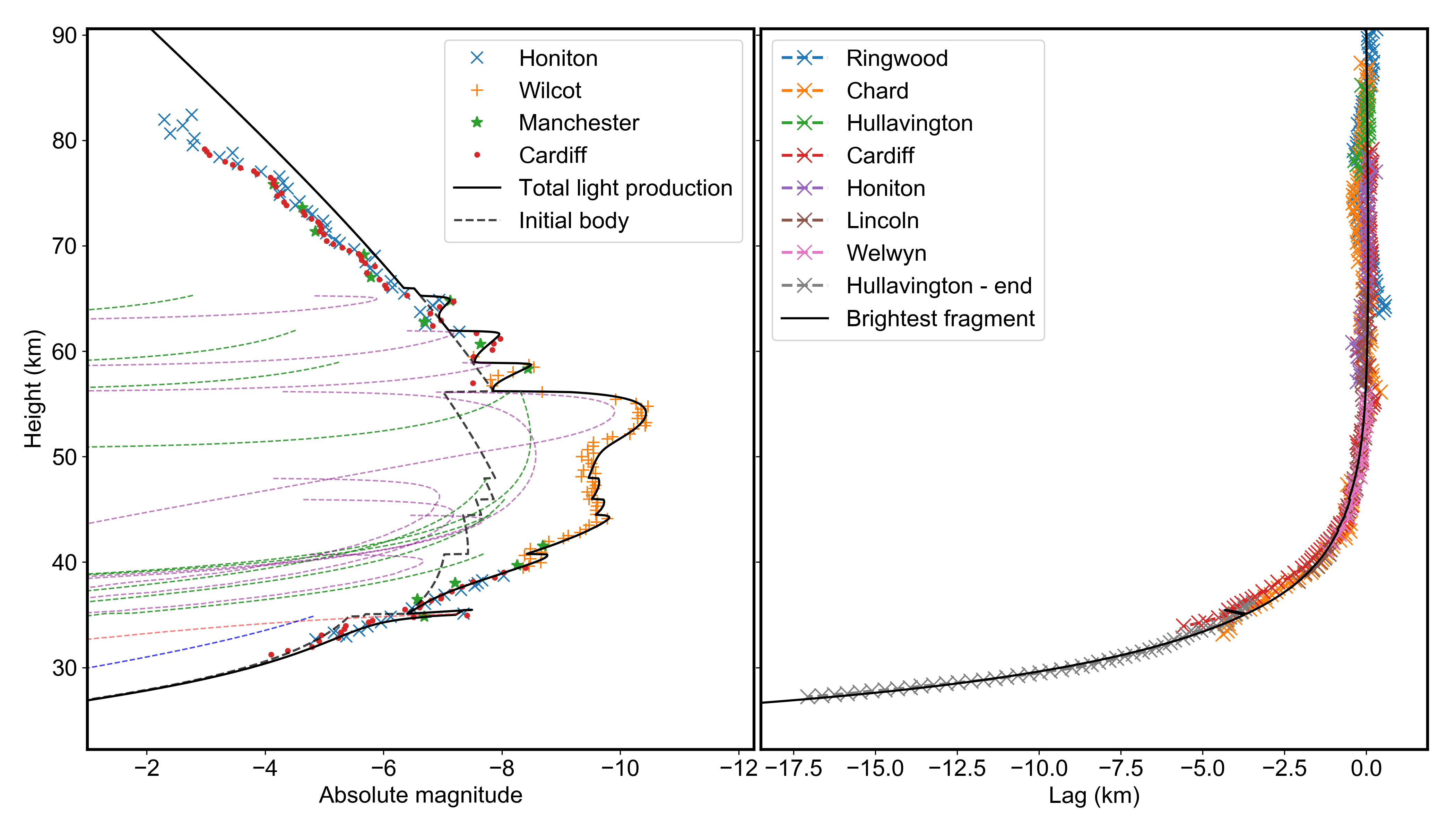}
    \caption{Fit of the ablation model (black line) to the observed light curve (left) and the lag (right) for the Winchcombe fireball. The lag is the distance a decelerating meteoroid falls behind a hypothetical non-decelerating meteoroid moving at the initially observed velocity \citep{vida2021high}. Inferred contributions to the total light production of individual fragments are given in dashed lines; black for the magnitude of the main body, green for eroding fragments, purple for grains released by erosion, blue for single-body fragments, and orange for dust released directly from the main body.}
    \label{fig:ablation_deceleration_model}
\end{figure}

An excellent fit to the data was obtained, within observational uncertainty (Figure \ref{fig:ablation_deceleration_model}). The only significant discrepancy between the model and observations is at the beginning of the fireball above $\sim60$~km. For this portion, the fit of the model to the light curve could not be improved, even with an unphysical and extremely low ablation coefficient of $\sigma = 0.0001$~\ablaunit{} ($50\times$ lower than the intrinsic $\sigma$ of ordinary chondrites). This suggests a significant influence of preheating of carbonaceous material, as was similarly noticed for ordinary chondrites \citep{spurny2020vzvdar}. The model indicates that the mass loss was not important in this initial stage, meaning the light production was purely caused by the drag component. This discrepancy calls for a revision of luminous efficiency models or a separation of models into mass loss and drag components with different values. For example, \cite{borovicka2011photographic} have measured a luminous efficiency of the drag component of the Hayabusa capsule re-entry, as its ablation shield prevented any significant mass loss. The capsule entered the atmosphere at 12~\kms{}, a similar speed to Winchcombe, and its observed luminous efficiency was 1.3\%. The luminous efficiency in our model was $\sim5\%$, which can explain the observed discrepancy. Nevertheless, the initial portion of the light curve does not play a significant role in deriving physical properties nor fragmentation behaviour and does not affect the conclusions below which stand for the given model assumptions.

The inverted physical properties of the fireball are given in Table \ref{tab:model_phys} and the fragmentation behaviour is given in Table \ref{tab:fragmentation}. We note that error estimation remains a challenge with this empirical method, and this might not be a unique solution, thus concrete error estimates are hard to give \citep{vida2022direct}. Nevertheless, the data is accurate enough to constrain the model bulk density to $\pm10$~\densunitSI{} while keeping all other parameters fixed. Progressive fragmentation was not modelled, so it was assumed that the released fragments do not have an intrinsic ablation coefficient $\sigma = 0.005$~\ablaunit{} of ordinary chondrites, but an apparent ablation coefficient appropriate for C-type meteoroids $\sigma = 0.04$~\ablaunit{} \citep{ceplecha1998meteor}.

\begin{table*}
\caption{Modelled physical properties of the fireball.}
\centering
\begin{tabular}{llrl}
Description & & Value\\
\hline\hline % inserts double horizontal lines
Initial mass (kg)                              & $m_0$    & 12.5  \\
Initial speed at 180~km (\kms{}) & $v_0$    & 13.86 \\
Zenith angle                                   & $Z_c$    & $48.08^{\circ}$ \\
Bulk density (\densunitSI{})            & $\rho$   & 2100  \\
Grain density (\densunitSI{})           & $\rho_g$   & 3000  \\
Ablation coefficient (\ablaunit{})   & $\sigma$ & 0.005  \\
(above 66~km)                                  & $\sigma$ & $<$0.0001  \\
Shape factor (sphere)                          & $A$      & 1.21   \\
Drag factor                               & $\Gamma$ & 0.8    \\
(above 66~km)                                  & $\Gamma$ & 1.0  \\
(below 35.1~km)                                & $\Gamma$ & 0.55    \\
\end{tabular}

\label{tab:model_phys}
\end{table*}

\begin{table*}
\caption{Modelled fragmentation behaviour. The fragment mass percentage in the table is reference to the mass of the main fragment at the moment of ejection. The mass distribution index for all grains was $s = 2.0$. The values of the dynamic pressure are computed using the drag factor $\Gamma = 1.0$.}
\centering
\begin{tabular}{lllllrrllllll}
Time$\mathrm{^{a}}$  & Height & Velocity               &  Dyn pres & Main $m$ & Fragment  & $m$ & $m$     & Erosion coeff            & Grain $m$  & Meteorites \\
(s)   & (km)   & (\kms{}) &  (MPa)    & (kg)       &       & (\%) & (kg)   & (\ablaunit{}) & range (kg) & Mass (kg)\\
\hline\hline % inserts double horizontal lines
 2.71 & 65.30 & 13.80 & 0.025 & 12.50 & EF &  0.2  & 0.025 & 1.50 & $10^{-9} - 10^{-8}$ & - \\
 3.06 & 62.00 & 13.78 & 0.040 & 12.45 & EF &  1.0  & 0.124 & 1.00 & $10^{-9} - 10^{-8}$ & - \\
 3.39 & 59.00 & 13.73 & 0.059 & 12.29 & EF &  1.5  & 0.184 & 1.00 & $10^{-9} - 10^{-8}$ & - \\
 3.68 & 56.25 & 13.68 & 0.083 & 12.06 & EF & 36.0  & 4.341 & 0.10 & $10^{-4} - 10^{-3}$ & - \\
 3.69 & 56.20 & 13.68 & 0.084 &  7.72 & EF & 46.0  & 3.550 & 0.80 & $10^{-5} - 10^{-3}$ & - \\
 4.58 & 48.00 & 13.22 & 0.214 &  4.04 & EF & 20.0  & 0.808 & 0.05 & $10^{-4} - 10^{-3}$ & - \\
 4.81 & 46.00 & 13.00 & 0.265 &  3.19 & EF & 30.0  & 0.956 & 0.05 & $10^{-4} - 10^{-3}$ & - \\
 4.98 & 44.50 & 12.78 & 0.307 &  2.20 & EF & 30.0  & 0.660 & 0.10 & $10^{-6} - 10^{-4}$ & - \\
 5.42 & 40.80 & 11.94 & 0.428 &  1.46 & EF & 37.0  & 0.541 & 0.05 & $10^{-4} - 10^{-3}$ & $10^{-3}$\\
 6.19 & 35.10 &  9.45 & 0.584 &  0.81 & D  & 28.0  & 0.226 & -    & $10^{-5} - 10^{-3}$ & \\
 6.23 & 34.90 &  9.35 & 0.591 &  0.58 & $3\times$ F  & 25.0  & 0.144 & -    & -                   & $3 \times 0.12$\\
  End$\mathrm{^{b}}$ & 25.31 &     - &     - &     - & -  &    -  &     - & -    & -                   & 0.35 \\
\hline
\multicolumn{10}{l}{$\mathrm{^{a}}$ Seconds after 2021-02-28 21:54:15.9 UTC.} \\
\multicolumn{10}{l}{$\mathrm{^{b}}$ Final mass of the main fragment at the end of ablation.} \\
\multicolumn{10}{l}{EF = New eroding fragment; D = Dust ejection; F = Single-body fragment.} \\
\end{tabular}

\label{tab:fragmentation}
\end{table*}

\begin{figure}
    \centering
    \includegraphics[width=1.\textwidth]{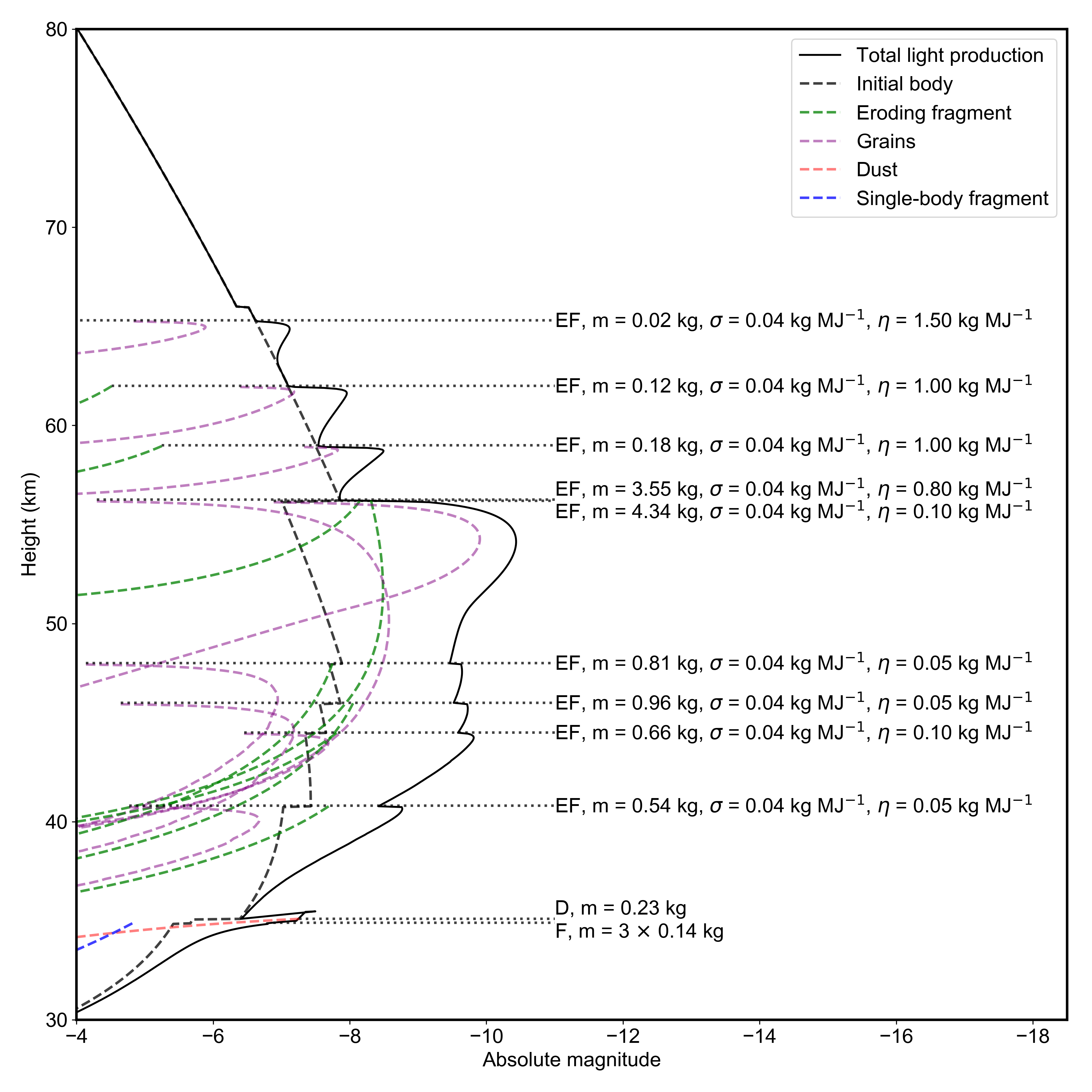}
    \caption{Details of individual fragmentation events. The fragmentation points, the type of fragmentation, total fragment mass, fragment ablation and erosion coefficients are shown.}
    \label{fig:model_details}
\end{figure}

\begin{figure}
    \centering
    \includegraphics[width=1.\textwidth]{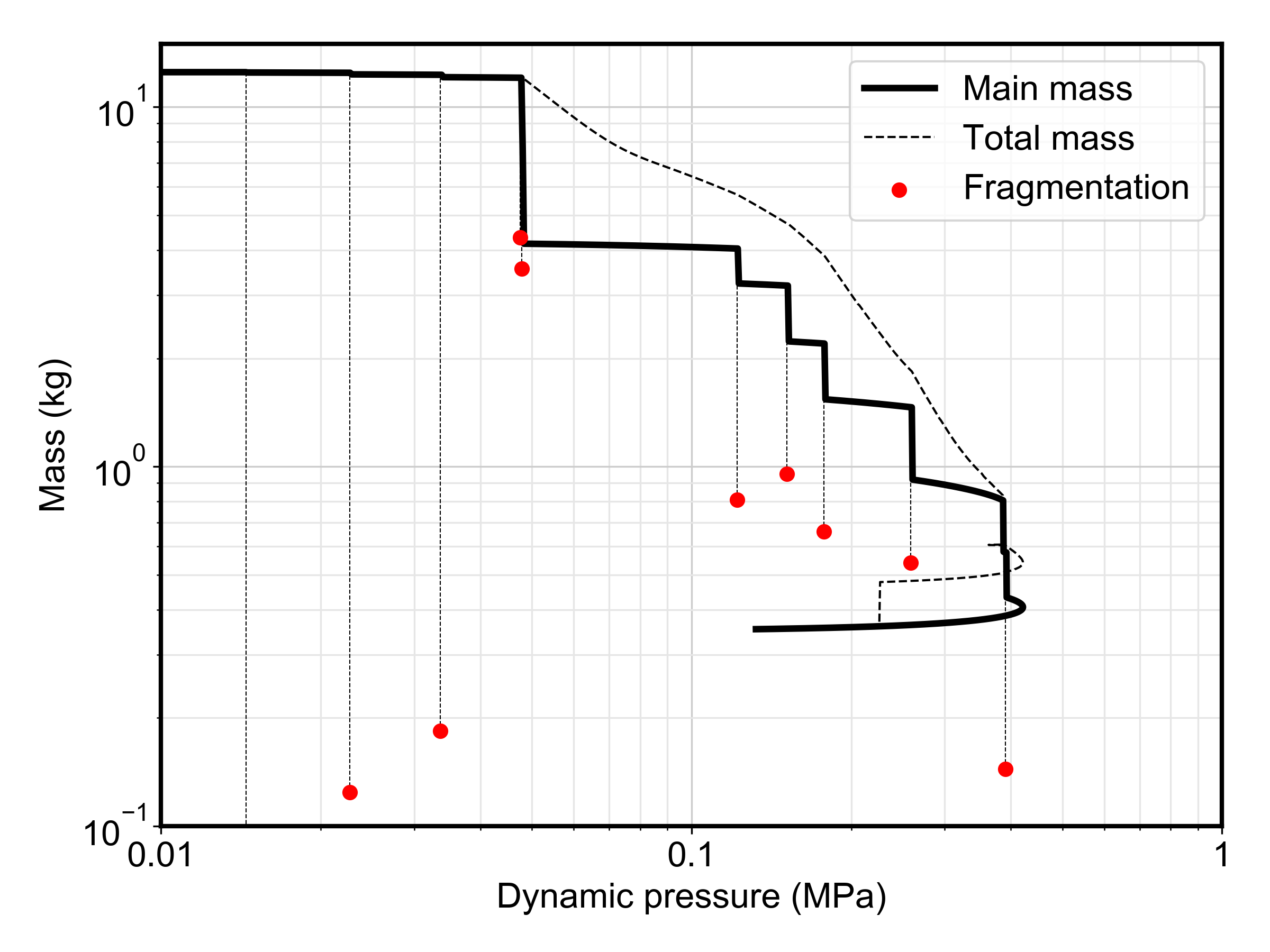}
    \caption{Mass of the main fragment (solid line) and the total mass (dashed line) versus the dynamic pressure. Individual fragmentation points and the total fragment masses are marked in red circles. The first fragmentation ($m = 0.02$~kg) is outside the plot for presentation purposes.}
    \label{fig:mass_loss}
\end{figure}

Figure \ref{fig:model_details} shows the details of the modelled fragmentation events. Starting with an initial mass of 12.53~kg, the meteoroid only experienced minor fragmentation (a loss of a few percent of mass) above 60~km of height, at dynamic pressures $P_{\mathrm{dyn}} \lesssim 0.06$~MPa (Figure \ref{fig:mass_loss}; $P_{\mathrm{dyn}} = \Gamma \rho_a v^2$, where $\rho_a$ is the atmospheric mass density at a given height and $v$ is the meteoroid speed at that height). However, the rapid mass loss occurred between $0.08$--$0.1$~MPa when over 80\% of mass was lost into two eroding fragments. The fireball continued to lose 20-30\% of instantaneous mass ($\sim0.5$~kg each) in each of the five subsequent fragmentations between $0.1$--$0.5$~MPa. Due to rapid deceleration, the rise in the dynamic pressure slowed and the fireball only reached a peak pressure of $0.6$~MPa at the height of 35~km. At this point of peak pressure, a final fragmentation with a sharp flare was observed which could only be explained by a sudden release of mm-sized dust.

Direct observations of the fragment train from the Hullavington station (Figure \ref{fig:frags}) showed that the final fragmentation event at the height of 35~km produced four distinct fragments. We reproduce the observed fragment mass distribution by assuming the final fragmentation produces three $\sim100$~g fragments. Even smaller fragments have also been observed in the video, but there is no practical way to directly measure their size. We note that 10 fragments total have been recovered \citep{russell_curation}, which is at odds with the four final fragments observed in the video. However, further fragmentation during the dark flight is a known phenomenon that might have been significant for this fragile body \citep{borovicka2003moravka, spurny2020vzvdar}, and some may even have been released higher up and obscured by the wake.

The fragmentation of Winchcombe occurred at dynamic pressures consistent with previously observed carbonaceous chondrite falls (Figure \ref{fig:mass_loss}), which have also all shown fragmentation in the $0.1$--$0.5$~MPa range \citep{borovivcka2019maribo, borovicka2021trajectory}. This is in stark contrast to ordinary chondrites which have two distinct phases of fragmentation, the first between 0.04--0.12~MPa and the second between 0.5--5~MPa \cite{borovivcka2020two}.

\section{Orbital analysis} \label{sec:orbit}

\subsection{Radiant and pre-encounter orbit}

The pre-encounter orbit was calculated from a trajectory that excluded all measurements below 35~km (Sec. \ref{sec:traj}), due to the fragment deviation below that altitude.
Orbital elements are given in Table \ref{tab:orbit}.
Winchcombe's orbit is well within the main belt (not evolved).
The semi-major axis (2.5855\,AU) places it between the 3:1 (2.5\,AU) and the 5:2 (2.82\,AU) mean-motion resonances with Jupiter, and points to these as probable mechanisms for delivering Winchcombe to near-Earth space.
Although Winchcombe shares a mid-belt semi-major axis with two other CM2 meteorites (Sutter's Mill \citep{2012Sci...338.1583J} and Maribo \citep{2019M&PS...54.1024B}), its Tisserand parameter with respect to Jupiter (T$_J$ = 3.12) places it on the asteroid side (T$_J >$3), contrary to the other two. The dynamical evolution of these objects can sometimes obfuscate their true origins \cite{shober2021main}.

\begin{table}[!h]
    \caption{Pre-encounter orbital parameters expressed in the heliocentric ecliptic frame (\textit{J2000}) and associated $1\sigma$ formal uncertainties.
}
    \centering
    \begin{tabular}{lcrlrr}
                           & Unit  & Value    & $1\sigma$     & \multicolumn{2}{c}{95\% Confidence Interval} \\
                           &       &          &               & Lower    & Upper \\
    \hline 
Semi-major axis            & AU    & 2.5855   &$\pm 0.0077$   & 2.5686   & 2.5980 \\
Eccentricity               &       & 0.6183   &$\pm 0.0011$   & 0.6158   & 0.6201 \\
Inclination                & \degr & 0.460    &$\pm 0.014$    & 0.440    & 0.490  \\
Argument of periapsis      & \degr & 351.798  &$\pm 0.018$    & 351.759  & 351.824 \\
Longitude ascending node   & \degr & 160.1955 &$\pm 0.0014$   & 160.1933 & 160.1985 \\
Perihelion                 & AU    & 0.986839 &$\pm 0.000012$ & 0.986814 & 0.986861 \\
Aphelion                   & AU    & 4.184    &$\pm 0.015$    & 4.150    & 4.209 \\
Tisserand's parameter      &       & 3.1207   &$\pm 0.0056$   & 3.1117   & 3.1331 \\
Last perihelion            & days  & 2021 Feb 22.446 &$\pm 0.015$ & 22.413 & 22.469 \\
Geocentric Right Ascension & \degr & 56.638   &$\pm 0.017$    & 56.604   & 56.671 \\
Geocentric Declination     & \degr & 17.713   &$\pm 0.069$    & 17.555   & 17.816 \\
Geocentric velocity        & \ms   & 8123     &$\pm 13$       & 8093     & 8143 \\
    \hline 
    \end{tabular}

    \label{tab:orbit}
\end{table}

\begin{figure}
    \centering
    \includegraphics[width=0.6\textwidth]{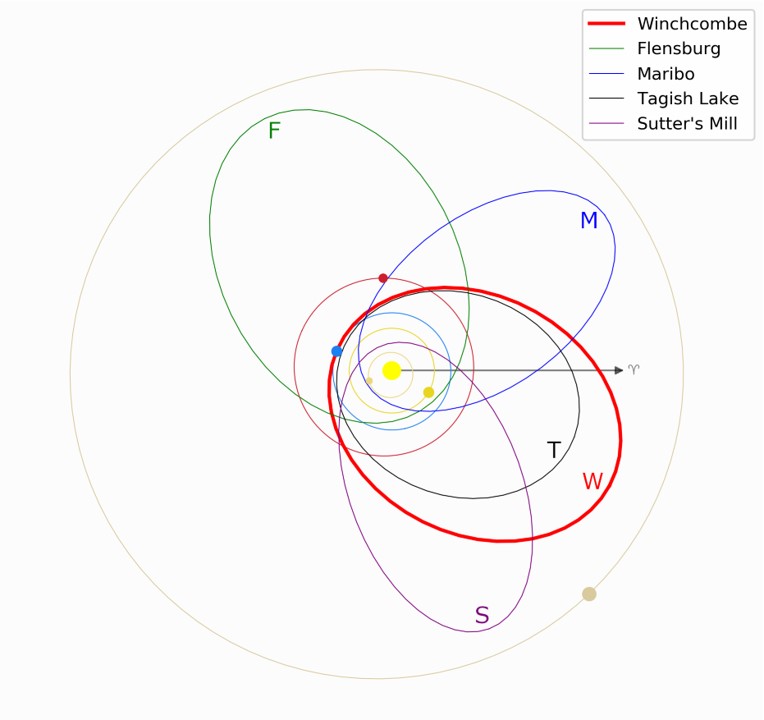}
    \caption{Ecliptic projection representing the pre-encounter orbit of the \meteorite{} meteoroid, in context with other carbonaceous orbital meteorites. Inclination to the ecliptic for \meteorite{} is 0.46\degr.}
    \label{fig:orbit}
\end{figure}

\subsection{Orbital history}

To gain insight into the recent dynamical past of the meteoroid before it crossed the Earth's path, we use backward integrations of the orbit following the method used by \citet{2021PSJ.....2...98S}.
1000 orbital clones of the meteoroid are created based on the uncertainties (Table \ref{tab:traj_sum}) and then integrated backwards using the Rebound IAS15 adaptive time step integrator \citep{2015MNRAS.446.1424R} with the Sun, 8 planets, and the Moon as active bodies.
The state vector of the test particles is recorded every 1000 years, both in barycentric coordinates and as osculating ecliptic orbital elements. Backward integrations ended at 3 million years in the past, way past the time for which meaningful dynamical insights into the meteoroid's history can be gained. Because the meteoroid was affected by Earth, Mars, and Jupiter in the recent past, the dynamical system is very rapidly chaotic. In post-analysis, we record at what time each meteoroid clone entered near-Earth space (perihelion distance $<1.3$ AU).
This gives us a median near-Earth entry of $\sim$0.08 Myr in the past, with 50\% of particles entering between 0.035 and 0.24 Myr ago.
In comparison, measured cosmic-ray exposure ages for Winchcombe are $\sim$0.3 Myr for $^{21}$Ne and 0.27 $\pm$ 0.08 Myr for $^{26}$Al \citep{king_science_winchcombe}.
This indicates that the ejection of the Winchcombe meteoroid from a larger parent asteroid (start of exposure to cosmic rays), and its orbital migration from the main belt to near-Earth space were either contemporaneous events or the ejection happened while the parent body was already in near-Earth space.

\begin{figure}
    \centering
    \includegraphics[width=0.5\textwidth]{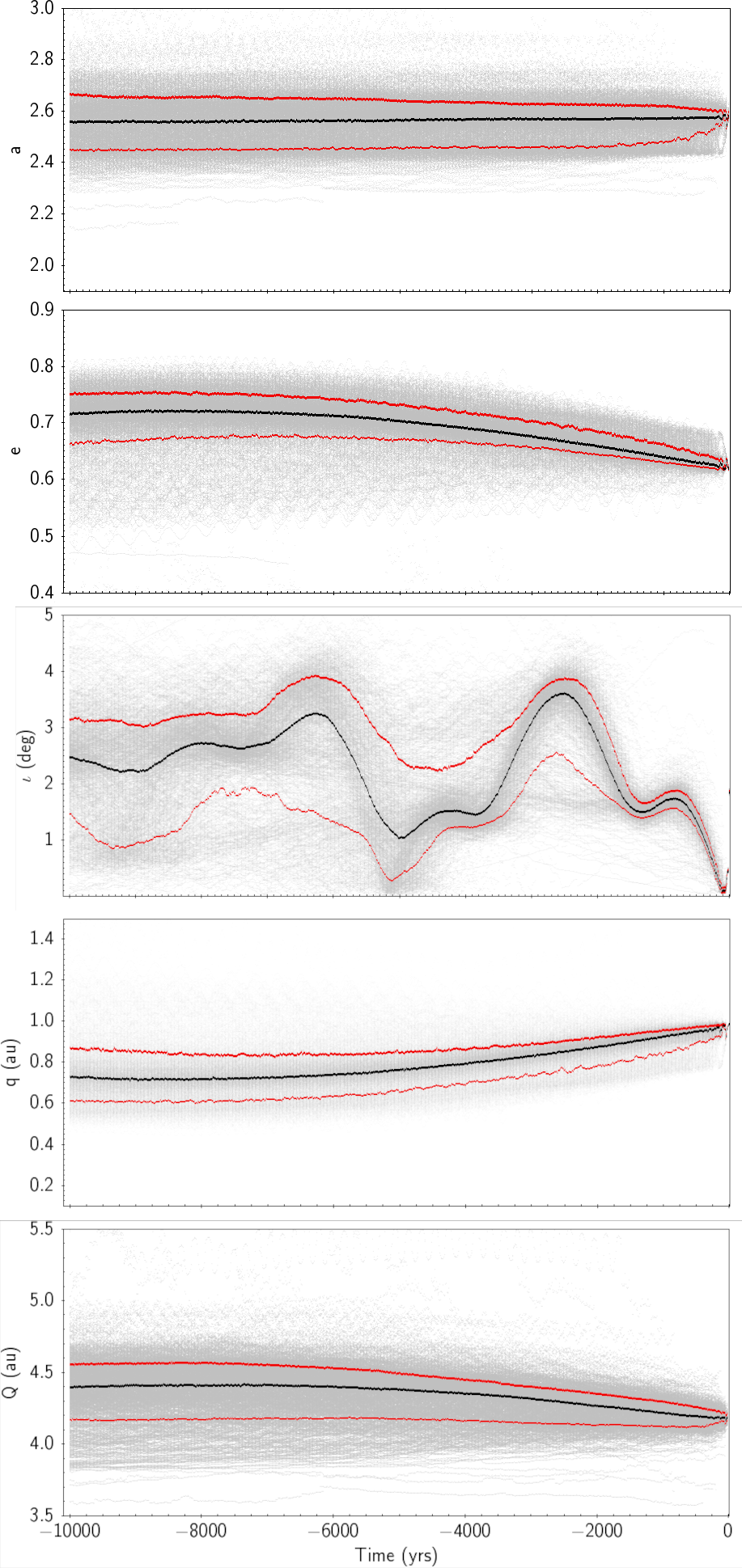}
    \caption{Orbital elements for the Winchcombe meteoroid over the previous ten thousand years, from the integration of 1000 particles using the IAS15 integrator \citep{2015MNRAS.446.1424R}. The black line is indicative of the median value, meanwhile, the red lines show the 1-sigma variation.}
    \label{fig:orbit_time}
\end{figure}

We also track the median perihelion distance of the particles: although the meteoroid has likely spent time closer to the sun than its impact perihelion distance suggests ($\sim$0.9868 AU), it most likely has remained higher (Figure~\ref{fig:orbit_time}) than that of both Sutter's Mill and Maribo \citep[$\sim 0.5$ AU; ][]{2021MNRAS.tmp.1743T}.
This suggests that in its recent NEO history, Winchcombe underwent less radiant heating ($\lesssim $400 K using \citet{2009MNRAS.400..147M}) than its orbital CM counterparts, which were heated to at least 100~K higher temperatures.

\section{Strewn field} \label{sec:darkflight}

\subsection{Atmospheric model}

\begin{figure}
% HARD done and checked 2021-03-18
    \centering
    \includegraphics[width=0.6\textwidth]{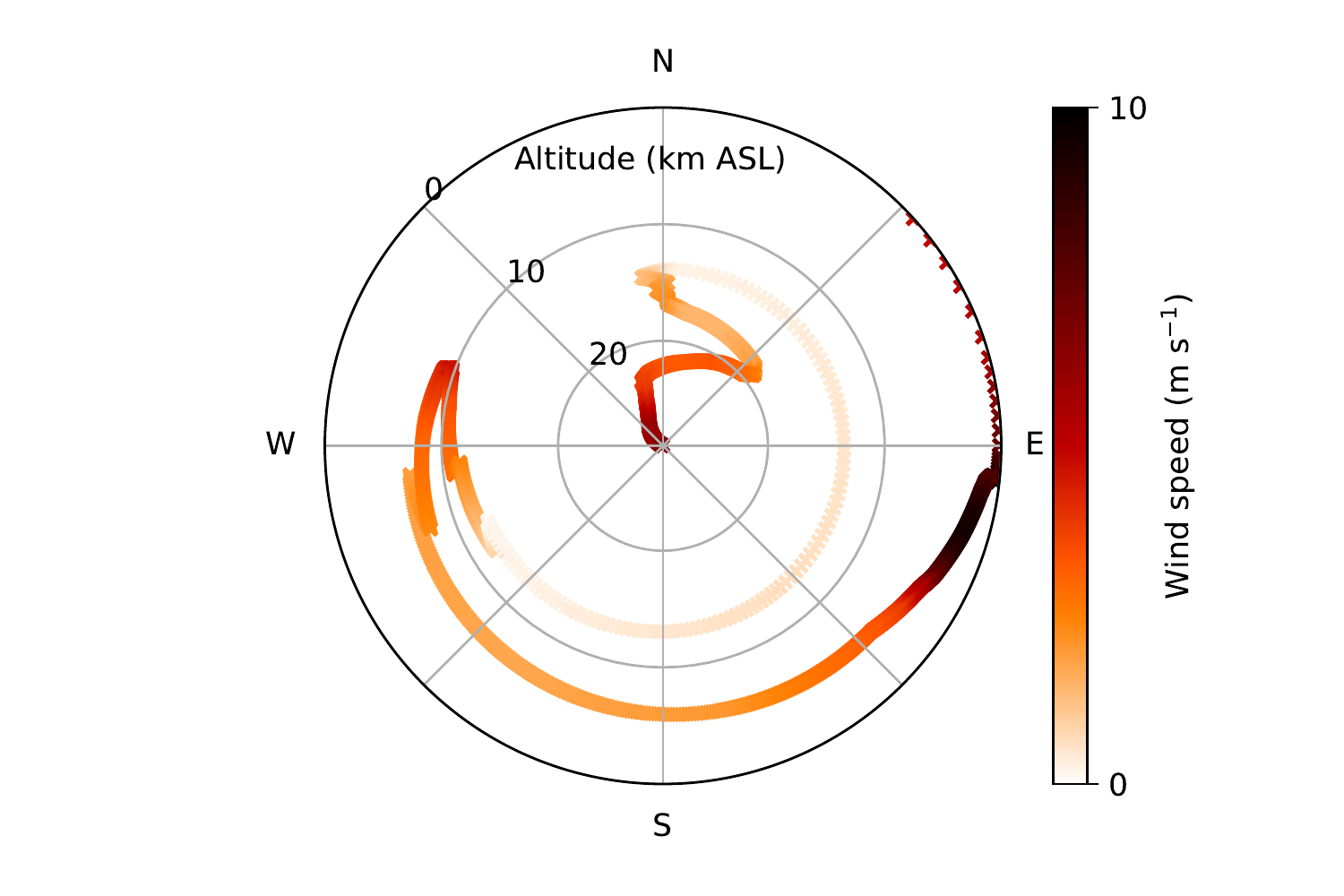}
    \caption{Wind model (speed and direction for a given altitude), extracted as a vertical profile at the location and time of the lowest visible bright flight measurement. Model integration started at 2021-02-12T12:00. The winds affecting this fall were very low, as the maximum wind encountered by the meteoroid during the dark flight was $\sim$10~\ms{} coming from the East, at around 0.4~km altitude.}
    \label{fig:Wind}
\end{figure}

The atmospheric conditions and winds for the dark flight were modelled numerically using the Weather Research and Forecasting (WRF) model version 4.0 with dynamic solver ARW (Advanced Research WRF) \citep{skamarock2019WRF4}.
The weather models include wind speed, wind direction, pressure, temperature, and relative humidity at heights ranging up to 30~km.
Three runs were processed, starting the weather simulation at different times before the meteorite fall, on 2021-02-28 at 6:00, 12:00, and 18:00 UTC. The 12:00 UTC profile is shown in Figure \ref{fig:Wind}. Fortuitously, the atmospheric conditions were stable and all three times gave relatively similar profiles, which is not always the case for other falls \citep[e.g. ][]{2018M&PS...53.2212D}. The atmospheric models are available in Supplementary Materials.

\subsection{Dark flight}

The 12:00 UTC wind model was used to predict where meteorites would land, based on the last observed bright flight state vector (Sec. \ref{sec:traj}), using the method of \citet{2022PSJ.....3...44T} \footnote{Code openly available at \url{ https://github.com/desertfireballnetwork/DFN_darkflight}}.
We create Monte Carlo clones by varying the final observed state vector within uncertainty (Sec. \ref{sec:traj}), as well as meteorite physical parameters such as mass, shape and density (fixed at 2090 \densunitSI{} based on recovered meteorite properties \cite{king_science_winchcombe}). The azimuth and altitude used for dark flight followed the original straight-line trajectory, prior to the deviation observed at 35~km, as the absolute amount and direction of the deviation could not be determined from single-station observations.
The masses are randomly sampled logarithmically from $\sim$5~g to $\sim$0.8~kg, and shape coefficients drawn from a normal distribution that covers predominantly spherical and cylindrical shapes: $A = \mathcal{N}(\mu=1.4,\,\sigma^{2}=0.33)$ 
% uniformly drawn from $A = 1.21$ (equivalent to a sphere) to $A =1.7$ (brick shaped) 
\citep{zhdan2007drag, 2017AJ....153...87S}.
The drag is calculated dynamically as a function of the atmosphere and flight conditions via the Reynolds number (see \cite{2022PSJ.....3...44T}).
For this case, a model was also run for extreme non-aerodynamic shapes (A=2.7; tile-like \cite{zhdan2007drag}) to extend the fall line to all recovered fragment locations (see discussion in section \ref{sec:along_line}).

\begin{figure}
    \centering
    \includegraphics[width=1.\textwidth]{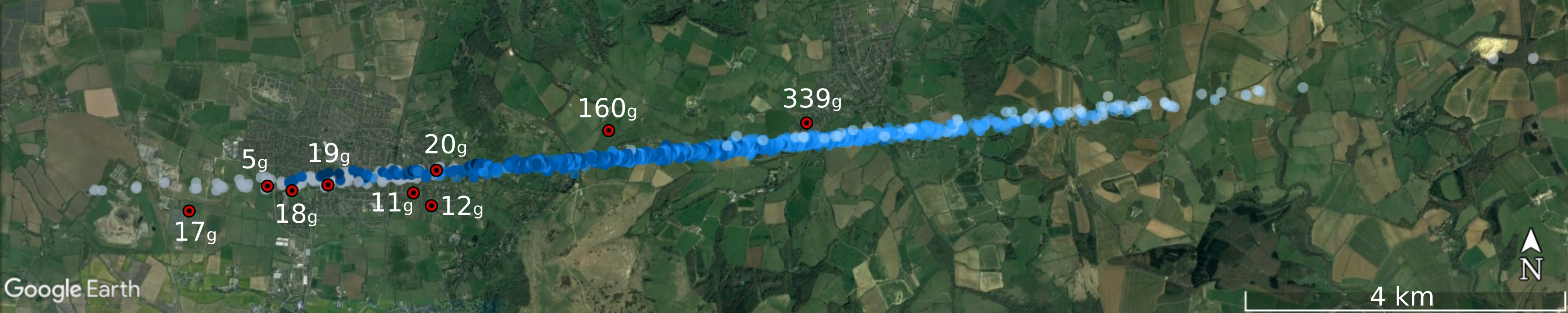}
    \caption{Fall line for the Winchcombe meteorite fall.
    Red points correspond to the location of the meteorites found with their estimated masses \citep{russell_curation}.
    Monte Carlo points from the darkflight simulation are blue shaded based on the assumed shape. In this case, the shape is the dominant parameter shifting the points along the fall line: dark blue are more brick-like and less aerodynamic in shape, while light blue is more sphere-like and aerodynamic shapes. Grey points to the west are from a custom model using an extreme, tile-like shape for reference. }
    \label{fig:fall_line}
\end{figure}

The predicted strewn field is in reasonable agreement with the positions of the recovered meteorites (Fig. \ref{fig:fall_line}).
Nonetheless, our modelling is unable to predict the lateral extent of the fall area. Below, we treat along and across the fall line shifts separately.

\subsubsection{Along the fall line}\label{sec:along_line}
The position of masses along the fall line is primarily controlled by the shape and the mass of the fragments (East-West direction in Fig. \ref{fig:fall_line}).
When performing Monte Carlo dark flight simulations, these parameters are assumed \emph{a priori} in order to predict where certain masses would fall (points in Fig. \ref{fig:fall_line}). 
Knowing the location and masses of recovered fragments with respect to these simulated stones can inform us of the approximate shape, impact speed, and flight time for each sample.

Cross-matching samples with simulations, the main mass found (Site 1 in \citet{russell_curation}) would have had a moderate shape coefficient of 1.4 to 1.5 (similar drag profile to a cylinder) and hit the ground at 46 to 49~\ms{} after a dark flight free fall time of $\sim$4.5 minutes. The other large mass (160~g at site 5) likely had a shape coefficient of 1.5 to 1.7 and hit the ground at 35 to 40~\ms{} after a flight time of $\sim$6 minutes.
Smaller fragments would require extreme shapes to match the fall line (grey points in Fig. \ref{fig:fall_line}). In particular, the 16.5~g mass at Site 6 requires a shape coefficient of $\sim2.7$ \citep[as found by][]{capek2021shapes} and would have impacted at 17 to 19 \ms, after a $\sim$16 minute flight. An alternative explanation is that this fragment entered the dark flight phase at a higher altitude (consistent with 2-3~km before the end), as was in fact observed in the Hullavington video. The lateral deviation of this fragment from the line also supports a longer dark flight phase (see also section \ref{sec:across_line}).
The spread of recovered meteorites along the modelled fall line is consistent with differences in their physical characteristics, as well as the late fragmentation of the smaller stones.

\subsubsection{Across the fall line}\label{sec:across_line}
The position of masses across the fall line is primarily controlled by the state vector at the end of the bright flight (North-South direction in Fig. \ref{fig:fall_line}).
The lateral spread of the Monte Carlo simulations due to uncertainties in this final state vector is $\sim$200~m (90~\ms{}). Some of the Winchcombe fragments were recovered outside of this area by up to $\sim$300~m on either side. This creates a $\sim$600~m corridor formed by the found masses.
It should be noted that our $\sim$200~m wide predicted fall line is in the middle of this $\sim$600~m corridor.  
To account for the fragments lying outside the lateral spread of the Monte Carlo simulations, the various meteoritic fragments must have had differences in their final bright flight velocity vector that model uncertainties alone cannot account for. 
Strong evidence of this direction change process is evident in Fig. \ref{fig:traj_residuals}, in which the Hullavington viewpoint displays a projected deviation of $\sim$200~m.
With a single viewpoint on the end of the bright flight trajectory, it is difficult to estimate the true spread of the fragments in 3D space.
Nonetheless, an observed 200~m projected deviation could well explain a 500~m cross-line positional difference between different fragments if their flight was completely tracked to the ground.

\citet{1980Icar...42..211P} have also proposed significant lateral velocities in fragments just after a breakup from studying strewn fields on Earth. These could be due to lift effects, bow shock interactions, centripetal separation of a rotating meteoroid, or transverse separation from reaching meteoroid crushing strengths \citep{1980Icar...42..211P}. 
These authors propose a spreading velocity proportional to the velocity at breakup such that $V_{spreading} = V_{breakup} \times \sqrt{C \times \rho_a /  \rho_m}$, where here the atmospheric density and meteoroid density refer to those at breakup altitude, and $C$ is an empirical constant that ranges from 0.02 to 1.5 \citep[based on strewn field analyses on Earth, Venus, and Mars; ][]{1980Icar...42..211P, herrick_effects_1994, popova_crater_2007, collins_meteoroid_2022}. For the final fragmentation of the Winchcombe meteoroid at 34.9~km, this relation predicts spreading velocities in the range of 2.5 - 22.5~m/s. This would account for a maximum deviation over the final 2.01~s of $\sim$45~m.
The observed 200~m deviation and a minimum lateral velocity of 90~m/s would require $C$ values of over 30. Such high lateral spreading velocities are not however unreasonable. Both \cite{borovicka2003moravka} for the Morávka fireball, and \citet{docobo1999video} for another fragmenting event, show spreading velocities of up to 300~m/s for observed fragments. 
These significantly higher values than represented in strewn field data show the energetic nature of the fragmentation processes that simple models of mere atmospheric loading cannot account for. The addition of any volatile materials in a carbonaceous-type meteoroid would be expected to exaggerate these effects. 
We note that the deviation occurred during a fragmentation event which released 28\% of the remaining mass ($\sim150$~g) into dust. We postulate that a preferential direction for the dust release might also explain the gain in transverse momentum.

%Note on spreading angle from Ellie:
%A common assumption for the rate at which equal-sized fragments separate after fragmentation is given by:
%v_spreading = vel * sqrt(C * (rho_a / rho_met)), note For winchcomb : 3 to 24 m/s!

%(Passey and Melosh, 1980, Artemieva et al., 2001; Popova et al., 2007), where rho_a/rho_met is the ratio of atmospheric density at the fragmentation altitude to fragment density, V is the along-trajectory velocity at fragmentation and C is an empirical constant. Estimates based on terrestrial crater strewn fields puts C between 0.02 and 1.5 \citep{PasseyMelosh1980} 
% excerpt from Passey and Melosh 1980:
% "Several mechanisms for supplying a transverse velocity component to the meteoroid fragments include: the effect of transverse lift, centripetal separation from a rotating meteoroid, dynamical transverse separation resulting from the crushing breakup of the meteoroid, and the interaction of two or more bow shocks of the fragments just after breakup. "
% for Winchcombe this could be up to 5 km.... think you're within that!

A common assumption for fall area estimation is that the more precise the overall bright flight trajectory is, the smaller the search area will be.
Although this is likely true, there is a catch: the individual fragments must be well observed from multiple sites at the end of the bright flight.
Therefore, unless multi-station viewpoints are available at high-resolution (arc minute) and high sensitivity (down to magnitude 0$^{\mathrm{M}}$), it is not possible to predict the width of meteorite fall line within less than a couple of hundreds of meters, no matter how precisely the upper trajectory is determined.

\section{Discussion} \label{sec:discussion}

The limited number of carbonaceous chondrites with known pre-atmospheric orbits is largely due to their poor survivability \citep{ceplecha1976fireball}. Carbonaceous chondrites are weak, so they fragment and ablate quickly in the Earth's atmosphere \citep{ceplecha1998meteor}. To survive as a meteorite, their atmospheric trajectories require one or more specific properties: low entry speeds (approach from the antapex, i.e. opposite to the direction of Earth's motion around the Sun), shallow entry angles, and large initial masses.

The first instrumentally observed carbonaceous chondrite meteorite fall, Tagish Lake, fell on 18 January 2000 at 16:43 UTC in a remote region of northern British Columbia, Canada \citep{2000Sci...290..320B}. The meteorite was categorised as a C2-ungrouped carbonaceous chondrite with an initial mass of up to 200,000~kg. There were over 70 eyewitness reports, with 24 photographs and five videos of the dust cloud taken by observers within 1-2 mins after the event. It was also detected by infrared and optical sensors aboard US Department of Defense Satellites and the corresponding shock wave was detected by local seismic and infrasound stations \citep{2002M&PS...37..661B}. More than 500 meteorites were recovered, with the largest fragment 2.3~kg, and a total mass of 16.3~kg \citep{2011M&PS...46.1525P}. The lack of direct ground-based optical recordings and reliance on classified satellite data mean that the orbit might contain systematic uncertainties that are not well understood \citep{2019MNRAS.483.5166D}.

The Maribo meteorite fell on 17 January 2009, at 18:08:28 UTC in Denmark \citep{haack2012maribo}, and was identified as a CM2 meteorite with a calculated initial meteoroid mass of 2000$\pm$1000~kg \citep{2019M&PS...54.1024B}. There were 550 eyewitness reports and it was captured by a surveillance camera in southern Sweden, a photo from an all-sky fireball camera in the Netherlands, and by three all-sky meteor radars in Germany. The sonic boom was recorded by 11 seismometers and an infrasound station. Seven radiometers in the Czech Republic were able to measure the radiometric light curve, allowing for the first detailed ablation and fragmentation modelling of a carbonaceous meteorite fall.

The Sutter's Mill meteorite fell on 22 April 2012, at 14:51:12 UTC in California and was identified as a CM2 carbonaceous chondrite with a calculated initial meteoroid mass of 20,000--80,000~kg \citep{2012Sci...338.1583J}. It was observed by 3 Doppler weather radars, 2 infrasound stations, and 8 seismic stations, along with a set of three photographs from Nevada, and a few videos. 

The most recent recorded carbonaceous meteorite fall, Flensburg, occurred on 12 September 2019, 12:50 UTC over northern Germany. It was identified as a C1-ungrouped carbonaceous chondrite with a calculated initial meteoroid mass of 10,000--20,000~kg \citep{2021M&PS...56..425B}. There were 584 eyewitness reports and it was recorded by one AllSky6 camera and three dash cameras. A single meteorite of 24.5~g was recovered, and there was insufficient data to estimate the total mass or number of fragments.

Each of the four previous carbonaceous falls had high entry velocities ($>15$~\kms{}) and consequently experienced high dynamic pressures ($>1$~MPa) which caused mechanical disintegration of the weak bodies. The reason why any meteorites survived at all is that their pre-atmospheric masses were large ($>$1~t) and they had shallow entry angles ($<30^\circ$) (Table \ref{tab:carb_chondrites}), allowing few meteorites to survive by chance while $>>99$\% of the initial mass was destroyed. In contrast,  Winchcombe experienced the most favourable entry conditions possible which enabled a significant amount of material from the smallest ever observed carbonaceous meteoroid to survive to the ground. 

\begin{table*}
\centering
\begin{tabular}{ccccccccc}
\hline
Name & Classification & Max & Initial Mass& Initial & Entry  & Semi-major & $T_J$ & Ref. \\
 &  &  Pressure&  Range  &  Velocity &  Angle &  Axis &  &  \\
 &  & (MPa) & (kg) & (\kms{}) & ($^\circ$)  & (AU) & & \\
\hline 
Tagish Lake     & C2(ungrouped) & 2.2  & 50,000--200,000 & 15.8             & 16.5           & $2.1\pm0.2$       & 3.66 & 1  \\
Maribo          & CM2           & 5    & 1,000--3,000    & 28.3$\pm$0.3     & 31             & $2.43\pm0.12$     & $2.95\pm0.11$ & 2, 3 \\
Sutter's Mill   & CM2           & $>1$ & 20,000--80,000  & 28.6$\pm$0.6     & 26.3$\pm$0.5   & $2.59\pm0.35$     & $2.81\pm0.32$ & 4 \\
Flensburg       & C1(ungrouped) & 2    & 10,000--20,000  & 19.43$\pm$0.05   & 24.4           & $2.82\pm0.03$     & $2.89\pm0.02$ & 5 \\
Winchcombe      & CM2           & 0.6  & 9--15           & 13.547$\pm$0.008 & 41.92$\pm$0.03 & $2.586\pm0.008$   & $3.121\pm0.006$ & 6, 7 \\
\hline
\end{tabular}
\caption{Physical and orbital properties of previous carbonaceous chondrite falls, in comparison to Winchcombe. The entry angle is reference to the horizontal. See references for details on falls and their orbits: (1) \citet{2000Sci...290..320B}, (2) \citet{haack2012maribo}, (3) \citet{2019M&PS...54.1024B}, (4) \citet{2012Sci...338.1583J}, (5) \citet{2021M&PS...56..425B}, (6) \citet{king_science_winchcombe}, (7) this work.}
\label{tab:carb_chondrites}
\end{table*}

\section{Conclusions}

Winchcombe is the first meteorite recovered in the UK in 30 years, the first carbonaceous chondrite recovered in the UK, and the first instrumentally observed meteorite fall in the UK.

The main scientific takeaways of this work are:

\begin{itemize}

\item The Winchcombe meteoroid entered with favourable entry parameters (low velocity and entry angle) to avoid $>$1 MPa dynamic pressures; an uncommon range of conditions that were necessary for the survival of weak carbonaceous material. So far, amongst meteorites with measured orbits, only multi-ton objects have been proven to drop carbonaceous chondrites.
In the more traditional meteorite dropping size (decimetre), a strong velocity bias exists against the survival of carbonaceous material. Winchcombe is the first evidence that survival of smaller cabronaceous bodies is possible, but only for objects approaching from the antapex which by rule have the slowest entry velocities.

\item The surviving fragments experienced a significant flight vector change before entering dark flight, obtaining a velocity kick perpendicular to a straight-line trajectory of at least 90~\ms{}. The physical phenomenon that caused this could not be definitely determined, but we investigate several possibilities that require further study. The deviation resulted in fragments being scattered beyond the nominal error boundaries of the fall area. Without detailed observations at both high-resolution and high sensitivity of the very end of the bright flight, it may not be possible to predict meteorite fall positions to better than a few hundred metres.

\item The recent ejection of Winchcombe from a parent asteroid as measured by the cosmic-ray exposure ages ($\sim0.3$~Myr) is longer or contemporary with the time spent as a Near-Earth Object inferred from its orbit (0.035---0.24~Myr, 0.08~Myr nominal). This indicates a minimal time delay between ejection from its parent body and the insertion into near-Earth space.

\item The analysis of the Winchcombe fireball involved five independent optical observation networks. This validates the need for standard data exchange procedures, as proposed by \citet{2020EPSC...14..856R}, in order to enable a quick turnaround time from the time the fireball happens to when a fall area is calculated.
\end{itemize}

\section{Supplementary materials}

Supplementary materials have been uploaded as a \textit{Zenodo} record at \url{http://doi.org/10.5281/zenodo.6685719}.
It contains fireball images (including calibration data), astrometry tables in \textit{Global Fireball Exchange} standard, the trajectory report file, and the wind profiles.

\begin{acknowledgements}

This publication is part of the Winchcombe science team consortium, organised by the UK Fireball Alliance and conducted by the UK Cosmochemistry Network. The authors of this paper would like to thank the UK Fireball Alliance, its constituent networks (UK Fireball Network, SCAMP, UKMON, NEMETODE, GMN), international collaborators (FRIPON, Global Fireball Observatory, Desert Fireball Network, University of Western Ontario, and University of Helsinki) and the meteor observation camera owners who participate in the UK Fireball Alliance network for their aid in observing the fireball and helping to predict its fall position. We would also like to thank the scientists and volunteers that participated in the UK Fireball Alliance led search and recovery of the Winchcombe meteorite, and the local community, who generously reported and donated meteorite finds and enabled the team to search the strewn field. STFC is acknowledged for supporting the “Curation and Preliminary Examination of the Winchcombe Carbonaceous Chondrite Fall” project (ST/V000799/1), and Natural History Museum staff for curatorial support.

The Curtin University authors acknowledge their contribution was made possible by the Australian Research Council as part of the Australian Discovery Project scheme (DP170102529, DP200102073), the Linkage Infrastructure, Equipment and Facilities scheme (LE170100106), and through institutional support from Curtin University. The Global Fireball Observatory data reduction is supported by resources provided by the Pawsey Supercomputing Centre with funding from the Australian Government and the Government of Western Australia, and makes intensive use of \textit{Astropy}, a community-developed core Python package for Astronomy \citep{2013A&A...558A..33A}.
The DFN gets software support resources awarded under the Astronomy Data and Computing Services (ADACS) Merit Allocation Program. ADACS is funded from the Astronomy National Collaborative Research Infrastructure Strategy (NCRIS) allocation provided by the Australian Government and managed by Astronomy Australia Limited (AAL).

This project has received funding from the European Union's Horizon 2020 research and innovation programme under the Marie Sklodowska-Curie grant agreement No 945298.

Several authors acknowledge funding from UK Science and Technology Facilities Council; SM and GSC ST/S0000615/1, LD ST/T002328/1, and KJ ST/R000751/1. AK is funded by UK Research and Innovation Grant MR/T020261/1. LD thanks a University of Glasgow COVID-19-Research Support Scheme grant.

\end{acknowledgements}

\bibliography{biblio}{}
\bibliographystyle{aasjournal}

%% This command is needed to show the entire author+affiliation list when
%% the collaboration and author truncation commands are used.  It has to
%% go at the end of the manuscript.
%\allauthors

%% Include this line if you are using the \added, \replaced, \deleted
%% commands to see a summary list of all changes at the end of the article.
%\listofchanges

\end{document}